# Reliable Postprocessing Improvement of van der Waals Heterostructures

Youngwook Kim[†, ‡, ∥], Patrick Herlinger[†, ∥], Takashi Taniguchi[§], Kenji Watanabe[§], and Jurgen H. Smet[†,*]

[†]*Max-Planck-Institut für Festkörperforschung, 70569 Stuttgart, Germany*

[‡]*Department of Emerging Materials Science, DGIST, 42988, Daegu, Korea*

[§] *National Institute for Materials Science, 1-1 Namiki, Tsukuba, 305-0044, Japan*

[∥]These authors contributed equally

[*]E-mail: j.smet@fkf.mpg.de

**Abstract**

The successful assembly of heterostructures consisting of several layers of different 2D materials in arbitrary order by exploiting van der Waals forces has truly been a game changer in the field of low dimensional physics. For instance, the encapsulation of graphene or $MoS_2$ between atomically flat hexagonal boron nitride (hBN) layers with strong affinity and graphitic gates that screen charge impurity disorder provided access to a plethora of interesting physical phenomena by drastically boosting the device quality. The encapsulation is accompanied by a self-cleansing effect at the interfaces. The otherwise predominant charged impurity disorder is minimized and random strain fluctuations ultimately constitute the main source of residual disorder. Despite these advances, the fabricated heterostructures still vary notably in their performance. While some achieve record mobilities, others only possess mediocre quality. Here, we report a reliable method to improve fully completed van der Waals heterostructure devices with a straightforward post-processing surface treatment based on thermal annealing and contact mode AFM. The impact is demonstrated by comparing magnetotransport measurements before and after the AFM treatment on one and the same device as well as on a larger set of treated and untreated devices to collect device statistics. Both the low temperature properties as well as the room temperature electrical characteristics, as relevant for applications, improve on average substantially. We surmise that the main beneficial effect arises from reducing nanometer scale corrugations at the interfaces, *i.e.* the detrimental impact of random strain fluctuations.

The study of two dimensional electron systems with extraordinarily low levels of disorder was for a long time the exclusive privilege of the epitaxial thin film research community. The successful isolation of graphene by mechanical exfoliation[1] and its dramatic quality improvements through suspension[2,3] or encapsulation[4] has however been truly disruptive. Even the most fragile ground states in two-dimensional electron physics have been unveiled in graphene. Also an unconventional superconducting ground state as well as fractional quantum Hall states at previously unknown filling factors were observed.[5-10] A plethora of other two dimensional materials has also been successfully isolated with rapid progress in sample quality as well.

The exfoliation technique appears the best route to obtain devices with very low intrinsic defect density and the performance of the 2D crystals is mainly limited by extrinsic disorder such as for example charged impurities located in the environment surrounding the 2D lattice. Intense efforts have been made to optimize the device fabrication and handling routines in order to reduce these extrinsic sources of disorder. The so-called dry pick-up method that makes use of the van der Waals forces between the layered materials, has become the key enabler for building arbitrary heterostructures of high quality for various reasons.[4] For one, by encapsulating the studied 2D material, for instance graphene or $MoS_2$, in between a van der Waals material with a high affinity to the active layer a self-cleansing effect occurs at the hetero-interfaces causing an aggregation of the contaminants into randomly distributed bubbles.[11] The remaining, bubble-free area of the heterostructure possesses atomically clean interfaces.[12] The bubble density itself can be reduced significantly by increasing the temperature during the stacking procedure[13,14] and subsequently performing annealing steps (see also SI Section 1).[15] The available area for designing high quality devices is then increased substantially. Hexagonal boron nitride (hBN) constitutes the candidate par excellence for encapsulation.[16] It combines a high affinity to graphene with a low intrinsic charged defect density.[11,17] The addition of graphitic gates in these van der Waals heterostructures also aids by screening charged impurity disorder in the substrate.[8,18] As charged impurity disorder is effectively minimized, another culprit takes over the scene. In a conclusive study, Couto and co-workers have demonstrated that random strain fluctuations make up the main source of residual disorder in already high quality graphene samples, *e.g.* on hBN.[19] In the low temperature limit, the mobility and the density inhomogeneity at low density, as estimated from the width of the field effect resistivity peak, are closely correlated. This finding can only be accounted for when random in-plane and out-of-plane deformations of the graphene lattice are the main source of long range disorder, which we elaborate in detail below. Hence, the atomic flatness of hBN[16] or other van der Waals layers supporting the active layer under study play a crucial role.

Despite all these measures, not all van der Waals heterostructures are created equal and a substantial variation among fabricated devices persists. Only a limited number does convincingly show the most fragile correlated ground states that can be anticipated in ultraclean 2D electron systems. Others only have mediocre quality and are usually not considered for further measurements. It is a matter of statistics. Achieving an improved yield for excellent quality devices therefore remains an important goal. In this Letter, we report a reliable method to boost the quality of such van der Waals heterostructures by means of a final step subsequent to the device processing and a high temperature annealing step. It is based on scanning an atomic force microscope (AFM) tip, operated in contact mode with a well-defined force, across the active device area preselected based on the absence of bubbles in optical microscopy and non-contact AFM. While the AFM tip efficiently cleans the top hBN surface from residues, which may act as a source of remote charged impurities, we surmise that the main beneficial effect arises from ironing away nanometer scale corrugations at the interfaces. This reduces the random strain fluctuations. Meanwhile we have applied this method to a multitude of fabricated devices and consistently obtain a higher yield of samples with enhanced properties, both statistically in terms of numbers as well as when making a comparison on one and the same device before and after the mechanical ironing procedure.

This Letter is organized as follows. We start with a detailed description of the heterostructure fabrication routine and the post-processing AFM treatment. To prove unambiguously its beneficial impact, low temperature magnetotransport measurements before and after the ironing are carried out on heterostructures consisting either of a graphene monolayer or a $MoS_2$ few layer, encapsulated in between thicker hBN flakes. The working hypothesis, that an ironing-like effect may occur, is further supported by manipulating and displacing larger AFM detectable bubbles with the tip as well as by a discussion of previous theoretical and experimental work addressing the different potential sources of disorder. We compare differences, advantages and disadvantages of our post-processing cleaning method with previous procedures put forward in the literature with the aim of reducing the bubble/blister free area and achieving improved quality. Finally, we provide some statistics on the performance of a larger set of devices as relevant for Hall sensing and thereby demonstrate that the ironing technique also enhances notably application relevant room temperature properties.

## Results/Discussion

### Device fabrication routines

All presented heterostructures were assembled with the dry van der Waals stacking method using a polymer stamp for the pick-up process and hBN flakes as encapsulating layers.[4,11] The polymer of the stamp was varied in order to increase the possible temperature range for the pick-up and release process as this range is set by the polymer's glass transition temperature. In general, a higher temperature improves the pick-up yield and also reduces the number of bubbles.[11,13,14] A poly-propylene carbonate (PPC)/PDMS, a commercial viscoelastic film (PF film from Gel-Pak), as well as an elvacite resin[20-23] were utilized. In cases where graphene is the active device layer, the stacks were annealed at 500°C in a forming gas ($Ar/H_2$) atmosphere after completed assembly and prior to further device processing. The purpose of the annealing is to lower the bubble density, promote the coalescence of residues and adsorbates in the remaining bubbles and increase the useable device area (see Section S1 in the supporting information and Fig. S1). We note that the stamp material did not affect the device quality nor the achievable improvement from the AFM treatment. However, the annealing step at 500°C turned out crucial to obtain a significant enhancement with AFM ironing (see Fig. S2). After this first annealing step, optical microscopy and non-contact AFM measurements were deployed to identify an area for the device design exhibiting neither bubbles nor wrinkles with these two methods. The device fabrication steps entailed etch mask and contact patterning using electron beam lithography with PMMA resist. Etching was performed with a $CHF_3/O_2$ mixture in an Oxford ICP machine.[22,23] Cr/Au edge contacts were thermally evaporated. A final annealing step at 350°C reduced the polymer residues from the device processing on the top of the hBN layer.

While the above processing routine was used for all graphene based van der Waals heterostructures, minor modifications were necessary for the $MoS_2$ devices. To avoid a degradation of the $MoS_2$ crystal, the annealing temperature after assembly was lowered from 500°C down to 350°C. Buried metal contacts (device M1) instead of edge contacts were chosen in order to obtain lower contact resistances. The buried contacts were achieved by etching the bottom hBN flake separately and depositing a metallic layer with a thickness approximately equal to the etch depth. The smaller the mismatch between the hBN and metal thickness the smaller the amount of strain induced in the $MoS_2$ layer. We refer to SI Section S3 for details on the recessed contact fabrication and device height profile. The device fabrication was then completed by releasing the pre-assembled and annealed van der Waals heterostructure consisting of a hBN/graphite/hBN/$MoS_2$ stack (from top to bottom) onto the hBN layer with buried contacts. The graphite layer served as a top gate. For all other devices presented in this work a doped silicon back gate was used instead to tune the carrier density.

Magnetotransport measurements, conducted on these finished devices prior to AFM cleaning, served as a reference in order to extract quantities that commonly serve as key figure of merits. These include for instance (1) the charge carrier mobility extracted from the zero field conductivity or the Hall coefficient,[24,25] (2) the onset of the Shubnikov-de Haas oscillations and appearance of quantum Hall plateaus,[24,25] as well as (3) the width of the charge neutrality peak recorded in a measurement of the conductivity *versus* charge carrier density.[19] The latter is a clear measure for the degree of the density inhomogeneity in the sample. In supporting information Section S4, Raman spectroscopy was also explored to assess inhomogeneity but approaches its limits to distinguish between different grades of high quality devices. The AFM cleaning was performed with a commercial system (NX-10 from Park Systems) operating in contact mode with a tip force between 50 and 150 nN and a scan speed between 0.3 and 0.6 Hz (maximum tip velocity of 6 µm/s). The mounted AFM tip is normally intended for non-contact mode measurements (Park PPP-NCHR) and has a guaranteed tip radius of curvature smaller than 10 nm and a tip height between 10 and 15 µm. For scan areas of up to 10 µm × 10 µm the line spacing was fixed to 512 corresponding to a maximum step size of 20 nm. 5 µm was the typical dimension of the cleaned areas. For the occasional cleaning of areas larger than 10 µm × 10 µm a line spacing of 1024 was chosen instead. The entire device channel was treated with the tip. The scan direction is not relevant and it was sufficient to perform one single contact mode scan to achieve the results demonstrated in this work. Non-contact AFM measurements were done before and after contact mode cleaning.

Experimental findings and discussion

Low temperature magnetotransport data recorded on a graphene stack (device G1) prior to tip cleaning is plotted in red color in panels a through c of Fig. 1. The curves reveal poor transport quality. A broad and asymmetric charge neutrality peak is observed and the mobility is only about 40,000 $cm^2/Vs$. The longitudinal resistance does not drop to zero in the quantum Hall regime, quantum oscillations are poorly visible and the plateaus in the Hall resistance are only reasonably developed for holes. Typically, one would sort out this device and fabricate another one. Instead, the sample was treated with the AFM tip and afterwards the transport measurements were repeated (black curves in Fig. 1). The width of the Dirac peak in the resistivity trace has shrunk substantially. Its position is now close to $V_g = 0$ V and the mobility is significantly enhanced to 350,000 $cm^2/Vs$. In the quantum Hall regime, the longitudinal resistance shows well developed quantum oscillations and nearly approaches zero. Hall plateaus are better developed and more numerous even at magnetic fields down to 1 T. We note that in the example shown in Fig. 1a the charge neutral peak has shifted indicating a reduction of the overall doping. However, in the majority of samples the position of the charge neutrality peak remains unaltered, yet the sample still exhibits comparable quality improvements as those seen in the example of Fig. 1a.

This straightforward method to improve the magnetotransport properties is also effective on van der Waals heterostructures possessing an active layer other than graphene. Fig. 2 illustrates magnetotransport data recorded on an encapsulated few layer $MoS_2$ device (M1) at a density of $5 \times 10^{12}$ $cm^{-2}$. An optical image of the device is shown in the inset to panel a. The red and black traces have been acquired before and after AFM cleaning, respectively. Panel a shows the raw data across the entire magnetic field range, while an enlarged window focusing on the low magnetic field regime is displayed in panel b. Prior to AFM cleaning, Shubnikov-de Haas oscillations develop around $B = 2.5$ T and signatures of incompressible symmetry broken ground states due to degeneracy lifting are observed at much higher magnetic fields only. After AFM cleaning, the quantum oscillations are much better developed and the onset is shifted downwards to $B = 1.7$ T. Additional quantum oscillations due to symmetry breaking are visible from $B = 2.5$ T onwards. We emphasize that this substantial improvement of the electronic quality has been achieved solely by the post-processing AFM treatment.

In an attempt to identify possible reasons for the sample quality improvement, we have analyzed the AFM images recorded in non-contact mode before and after tip treatment. An example is illustrated in Fig. 3b. In the left image, the channel of this device was already free of bubbles and wrinkles before the AFM procedure, at least within the resolution of conventional AFM. Yet, contrary to expectation, the transport quality as discussed previously was poor. Some residues are visible on the top hBN surface. These residues were mechanically removed during the contact mode AFM scan as seen in the right image in Figure 3b. The surface of the top hBN layer becomes flatter as highlighted in Figure 3b and d. This mechanical residue cleaning is however not the cause of the substantial improvement in device quality. In contrast to earlier reports on a substantial performance gain due to the mechanical cleaning of uncovered graphene,[26,27] the conducting 2D material under test here is completely sandwiched by thicker hBN layers. Therefore, the residues are not in direct contact and should not constitute a major source of disorder. Indeed, a former study on graphene based van der Waals heterostructures revealed that encapsulated graphene is essentially insensitive to polymer residues and the device performance was reported to be identical with and without polymer residues (see SI Section 1.1 and Fig. S2 in Ref. 4). A similar test has been performed here (see Section S5 in the SI). A sample was first treated by AFM ironing and studied. Subsequently, this device was intentionally contaminated with polymer residues by spin-coating of PMMA and subsequent standard cleaning. This device was then reinvestigated. The unavoidable residues of this additional processing step had no impact of any significance. The Full-Width-Half-Maximum of the charge neutrality peak was 55 mV and 52 mV before and after contamination. The gate voltage difference at charge neutrality was only 3 mV as shown in Fig S6. This corresponds to a density change at a given gate voltage of only $1\times10^9/cm^2$, below the accuracy with which the absolute value of the density can be determined. Other investigated devices already appeared clean on the top surface after the annealing step and did not show signs of the accumulation of residues on top of the hBN cap layer. Yet, they still consistently benefited from the AFM treatment. Residues may have been removed or burned away during the annealing procedure at high temperature. For the sake of completeness we note that it has also been reported that annealing can spread contaminants into a thin film.[28]

The $MoS_2$ sample addressed in Fig. 2 was fabricated with a 15 nm thick multilayer graphene sheet on top and capped by an additional hBN layer. The screening length for a perpendicular electric field of the graphite thin film, that served as a gate, has been estimated to be on the order of 1 nm, well below the thickness of the incorporated layer.[29] It therefore efficiently screens charged impurity potentials that may be generated by residues on the top sample surface. Such residues and their subsequent removal during AFM treatment can therefore not be the root source of the initial poor quality and the quality improvement after treatment. All of these observations provide strong evidence that the removal of residuals on the top surface are not a viable explanation for the observed improvement in the sample properties.

Our working hypothesis is that the AFM tip is capable of mechanically manipulating the buried interfaces and abating local sub-nanometer corrugations of these interfaces. This scenario would offer an immediate and plausible explanation for the observed improvement in the electrical transport characteristics of the AFM treated devices. Indeed, local strain fluctuations have in the past been spotted as an important remaining source of long range extrinsic disorder limiting the quality in otherwise clean van der Waals heterostructures.[19] These random strain fluctuations then constitute both the dominant carrier scattering mechanism as well as the main cause of spatial density inhomogeneity near the CNP, *i.e.* residual total carrier density (see also the theoretical works in Refs. 30,31 describing electron hole puddle formation due to strain fluctuations and the detailed discussion about different sources of disorder in SI Section S6). As it is central to our results, we summarize the key observations as well as the chain of arguments that led the authors of Ref. 19 in their systematic and careful study to this conclusion. When performing experiments on a large set of samples supported by various substrates, a striking inverse correlation was identified between the charge carrier mobility and the residual charge density as extracted from a log-log plot of the conductivity *versus* density. This correlation indicated that mobility and residual density have a common microscopic origin. From an analysis of the weak localization effect, intervalley scattering, requiring a large transfer of momentum, was found to be orders of magnitude longer than the transport scattering time from the mobility. Hence, not short range, but long range disorder limits the mobility. Long range disorder can either originate from charged impurities and residues or from local strain fluctuations. However, charged impurities only generate a scalar long range disorder potential. Such a potential is not capable of causing backward scattering in the case of graphene, since states of opposite momentum within a valley are orthogonal as they belong to the different Bloch bands of the two sub-lattices. A scalar potential can only cause intravalley scattering, if it is short range. This leaves only local strain fluctuations as the possible source of disorder which we explain in greater detail in SI Section S6. They cause not only a long range scalar potential, responsible for a fluctuating local charge density and electron-hole puddles, but also a spatially varying effective vector potential that does allow for intravalley backscattering despite its long range nature.[19,30,31] Consistency between the observed time scales for intervalley, intravalley and transport scattering was only achieved when accepting that the dominant source of the disorder and the residual charge density are local strain fluctuations. If the act of scanning the AFM tip across the top surface of the heterostructure is equivalent to ironing out sub-nanometer local roughness, local strain fluctuations would be mitigated and the transport quality improved. Hence, the conclusions of Couto *et al.*[19] are consistent with our experimental observations as well as the previous reports that residues on top of hBN have no impact on device quality.[4] The latter has been verified in our devices in Section S5 of SI.

In order to demonstrate that mechanical manipulation of interfacial roughness is in principle possible, we have applied the ironing procedure to other areas that exhibit larger, AFM detectable bubbles. The force needed to move those bubbles appears to scale with their lateral dimensions. An example of a heterostructure with three micrometer sized pockets, marked as A, B, and C, is shown in Figure 3e. After contact mode tip scanning, pocket A disappeared, *i.e.* likely ruptured, while B and C moved in the direction of the scan. However, we find the success rate of moving micrometer size bubbles by means of an AFM tip to be very low. The majority of the bubbles, larger than one micrometer in size, remain at their position even when forces up to several thousand nN are applied as shown in Figure 3e and f. The applied force more likely rips apart the bubble thereby damaging the heterostructure.[12] In this work, much smaller bubbles or interfacial roughness with sub-nanometer thickness are relevant and small forces on the order of 100 nN may be sufficient to manipulate and flatten the interfaces reliably. After all, other groups have demonstrated little friction among the different layers of van der Waals heterostructures, *e.g.* graphene and hBN, so that local relative motion may occur and be enforced. For instance, thermal annealing can induce a rotation and sliding of graphene between two hBN flakes.[32] Single constituents of a heterostructure have also been twisted *in situ* by an AFM tip.[33] While plausible, the flattening of sub-nanometer scale roughness and the accompanying release of in-plane strain fluctuations at the interfaces between the active 2D material and the hBN protective layer can unfortunately not be quantified directly, since conventional AFM does not allow observing sub-nanometer height changes at an interface buried by a hBN cap layer of at least a few nanometer. Other techniques to analyze the interfacial roughness are either destructive, restricted to areas of a few nanometer only or are insufficiently sensitive (see also SI Section S4 demonstrating what Raman spectroscopy at room temperature can accomplish in this regard).[12,34] Even if so, it is possible to indirectly corroborate that the AFM treatment action is equivalent to ironing and local strain reduction. Following the procedure of Couto *et al.*[19] we estimated the residual total carrier density at the CNP $n_{t,0}$, comprising the disorder induced electron hole puddles at low temperature, by a graphical analysis in a double-logarithmic plot of the conductivity as a function of density (see Fig. S7). Subsequent to the AFM treatment, the residual density $n_{t,0}$ is indeed reduced substantially and the mobility μ is enhanced as well. The ratio of these two quantities follows $1/\mu = (h/e) * n_{t,0} * 0.118$ as predicted by the theoretical model of random strain fluctuations in the work of Couto *et al.*[19] Here, h is Planck's constant and e is the elementary charge. As mentioned previously, a prerequisite to achieve the quality improvement after AFM treatment is a preceding high temperature annealing step. Without the subsequent AFM treatment no sample improvement is achieved. At present we believe that the annealing is instrumental to mobilize any contaminants that may be present at the heterointerfaces, so they can merge and coalesce in the existing bubbles, while the subsequent AFM treatment releases the strain in the cleaned areas. Although we have not explored the thickness limit of the hBN capping the device up to which samples benefit from the AFM treatment, the AFM treatment procedure has been applied on samples with hBN layers with a thickness of up to 50 nm. Such devices still benefit from the AFM treatment. These hBN layer thicknesses are at the upper end of what is typically used in van der Waals heterostructures and therefore the relevant parameter space has been covered.

We note that Rosenberger *et al.*[35] have also used an AFM tip with the aim of improving sample quality through flattening. However, the procedure is different and the applicability of this method covers a different region of parameter space. The flattening is achieved by injecting a water/solvent mixture during the pick-up and transfer method. As a result, the transferred material is floating on top of any contaminants that are available on the supporting substrate underneath. Scanning of the AFM tip has the two-fold purpose of displacing the contaminants to the edge of the scanning area and of flattening the heterointerface. Annealing is detrimental and attempts to apply this method on a heterointerface 18 nm below the top surface indicated that contaminant displacement is seriously hampered. It requires forces of 1000 nN or more. In our work the amount of initial contaminants is minimal and a solvent/water mixture is avoided altogether as we have found it detrimental to achieve the ultimate in transport quality. If any contaminants are still present at the heterointerfaces they are mobilized and removed through annealing. The AFM treatment is responsible for flattening only. Both methods have their value set. For heterointerfaces buried by a thicker capping layer or for the assembly of a van der Waals stack without any exposure of the active layer and internal heterointerfaces to solvents as required for ultimate quantum transport properties the method by Rosenberger *et al.*[35] is unsuitable, but our method is applicable. Materials that degrade during annealing at high temperature can only benefit from the method of Rosenberger *et al.* This complementarity makes both methods valuable.

Work by Purdie *et al.*[14] is also relevant in this context. It aimed at increasing the bubble or blister free area in van der Waals heterostructures. This was achieved by exerting a force with a stamp at a small angle after assembly of the heterostructure and prior to electron beam lithography. The authors indeed succeeded in increasing the blister free area and hence in enlarging the useable device area. A before/after demonstration was not possible as this stamp method is no longer applicable once the device is completed or contacts are made. In the present work, we highlight a post-processing technique that is applicable after the device is entirely completed. Transport characteristics can be compared before and after AFM ironing. It is not intended to remove blisters, but rather to improve quality further in areas that were initially blister free, yet do not exhibit the highest degree of quality. In the course of our experiments, it also has been observed that devices occasionally degrade during operation, but that this degradation can be healed efficiently by repeating the AFM ironing technique. The method described by Purdie *et al.*[14] has been attempted in our laboratory, but was only partially successful in the sense of enlarging the blister free area. In essence, both techniques have their justification and presumably combining them one obtains the best of both worlds. The stamp method aids in maximizing the blister free area, despite our own limited success with this approach, while the AFM ironing method brings out the best transport quality after device fabrication from the selected blister free area to pattern the device. If there is no need to increase the blister free area, the AFM ironing method is sufficient, reliable and straightforward to achieve excellent quality.

**Improvements of the device performance at room temperature**

Until now, we have discussed the notable improvement of the electrical properties and the quantum transport features at low temperature. Finally, we turn our attention to the impact of AFM ironing on the room temperature transport properties of graphene as relevant for applications. Among these potential applications are Hall sensors, which have been intensively studied in recent years as encapsulated graphene may offer exceptional sensitivity.[36-39] Fig. 4 illustrates typical room temperature data recorded on a 500°C annealed and hBN-encapsulated graphene device (G2) before and after AFM treatment. Panel a displays the density dependence of the conductivity, $\sigma(n)$. The density interval around the CNP for which the conductivity remains at its minimum narrows considerably after AFM ironing. The rise of the conductivity as we move away from the CNP also becomes steeper indicating that the mobility, mostly limited by long range disorder at low density, increases. Empirically, the somewhat non-linear density dependence of the conductivity can be described by a density-independent mobility due to long range scattering, $\mu_{LR}$, and a density-independent resistivity related to short range scattering.[16,40] Applying this model to the measured curves yields $\mu_{LR}$ = 89,000 cm$^2$/Vs before and $\mu_{LR}$ = 167,000 cm$^2$/Vs after the AFM treatment. These values were averaged between the electron and hole side. This is a significant improvement by nearly a factor of 2 through the application of such a straightforward post-processing AFM treatment step. We stress that the key message of this paper is not that higher mobility values can be achieved than previously reported,[4,14] but rather that it is possible by a post-processing procedure to reach these values on a routine basis. At high densities, they correspond to the phonon limit (see also Fig. S8).

Panel b of Figure 4 displays the density dependence of the field effect mobility, $\mu_{FE} = \sigma / e\, n$, as well as of the Hall mobility, $\mu_H = R_H\, \sigma$, before and after AFM cleaning. Here, $R_H$ is the Hall coefficient. It has been obtained from measuring the transverse resistivity while sweeping the magnetic field in the classically weak regime (here +/- 50 mT) and is also displayed in panel c. Both expressions for the mobility assume that transport occurs *via* a single carrier type, which is not valid near the CNP where electrons and holes coexist. This is also apparent from the Hall coefficient. The maximum at negative voltage and the minimum at positive voltage demarcate the regime of coexisting electron and holes. In this regime, the field effect mobility diverges and the Hall mobility goes to zero at the CNP. The overall mobility enhancement subsequent to the AFM treatment is again clearly visible in panel b. The Hall coefficient surpasses 5000 Ω/T, the earlier reported record sensitivity.[38] Panel d in Fig. 4 summarizes the maximum Hall sensitivity achieved on AFM treated and encapsulated graphene devices in this work as well as devices not treated by AFM. Even though the data for the treated and untreated samples were not all recorded on the same set of devices, the Hall coefficient distribution lies entirely above that of untreated devices. This yields strong statistical evidence for a higher Hall sensitivity in AFM treated devices, while non-treated devices perform on average weaker. Elsewhere, it has been demonstrated that the maximum Hall coefficient values are closely correlated to the residual total carrier density at the CNP $n_{t,0}$, since $R_{H,max} = 1 / 2\, e\, n_{t,0}$.[37,39] For our best AFM ironed devices we find $n_{t,0} \sim 6 \times 10^{10}$ cm$^{-2}$. Similar values can also be estimated from the conventional graphical analysis of a double-logarithmic diagram plotting the conductivity *versus* density,[19,41] as illustrated in Figure S2. These achieved values, $R_{H,max}$ and $n_{t,0}$, constitute intrinsic performance limits of graphene based Hall sensors due to thermal excitation of charge carriers at room temperature (see also Section 8 of SI).[37,42] The post-processing AFM treatment apparently enables to reach this intrinsic limit reliably. Charge inhomogeneity is no longer relevant, but instead the simultaneous presence of electrons and holes due to thermal excitation.[41,42] We may anticipate that under these conditions scattering in the single carrier, high density regime is no longer dominated by extrinsic disorder, but rather acoustic phonon scattering. This is confirmed in additional resistivity data in the supporting information (Fig. S2 and S8).

## Conclusions

In summary, the electrical properties of van der Waals heterostructures, that are receptive to self-cleansing effects, can be substantially and reliably augmented by annealing and post-processing

AFM cleaning through the exertion of a constant force between 50 and 150 nN with the tip. Devices with initially poor performance can be healed. Not only low temperature, but also room temperature electrical characteristics, as relevant for some applications, are improved and frequently reach intrinsic limits set by the thermally excited charge carriers. Previous work has identified random strain fluctuations as the key remaining culprit for extrinsic disorder in otherwise clean and high quality heterostructures.[19] It is therefore plausible to surmise that the annealing mobilizes remnants at the heterointerface and promotes their coalescence into larger bubbles, while the AFM action should be thought of as ironing away small local corrugations at the heterointerfaces, thereby reducing the role of local strain fluctuations for transport scattering.

## Methods

Here, we provide further details, not already discussed in the main text, on the source of the 2D materials as well as the mechanical exfoliation protocol. The latter is different for graphene and $MoS_2$ based Van der Waals heterostructures. The natural graphite stems from NGS Naturgraphit and the $MoS_2$ was purchased from 2D Semiconductor. Mechanical exfoliation was performed with Nitto tape BT-150P-LC, as according to our judgement it leaves the minimal amount of residues. For the graphene structures, the silicon substrate with a 300 nm dry thermal $SiO_2$ cap layer was treated with an $O_2$ plasma immediately prior to exfoliation. This procedure enhances the flake size remaining on the substrate, but it comes at the expense of a dramatically reduced pick-up yield during the subsequent van der Waals stacking, if no further measures are taken. An annealing step at 500°C for 30 minutes in forming gas mitigates the issue and boosts the pick-up yield back to close to 100%. $MoS_2$ is directly cleaved onto a viscoelastic stamp (Gel-Pak, PF-30/17-X4) with the same Nitto tape. Suitable $MoS_2$-flakes are selected with optical microscope and transferred to the $SiO_2$ substrate. In order to clean any residues on top of $MoS_2$ from the Gel-Pak, we annealed the $MoS_2$ at 200°C for 2 hours in forming gas. Alternatively, the $MoS_2$-flakes can also be mechanically cleaned using contact mode AFM with an exerted force of 5 to 50 nN.

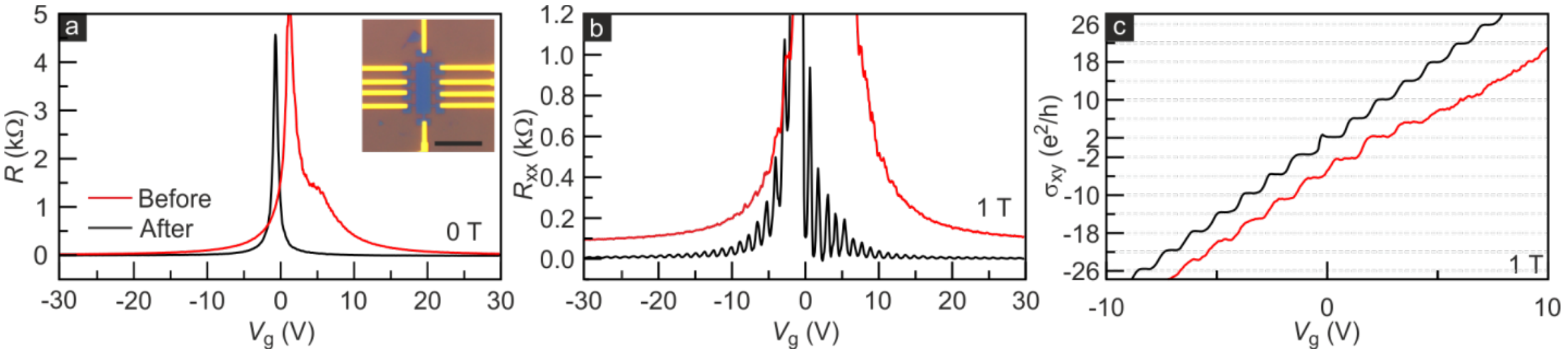

**Figure 1.** (a) Four terminal longitudinal resistance as a function of the gate voltage for device G1 in the absence of a magnetic field and $T$ = 1.3 K. The inset shows a microscopic image of the device. The scale bar corresponds to 10 μm. Longitudinal resistivity (b) and Hall conductivity (c) as a function of gate voltage at $B$ = 1 T and $T$ = 1.3 K. The red and black lines show the results before and after the AFM ironing, respectively.

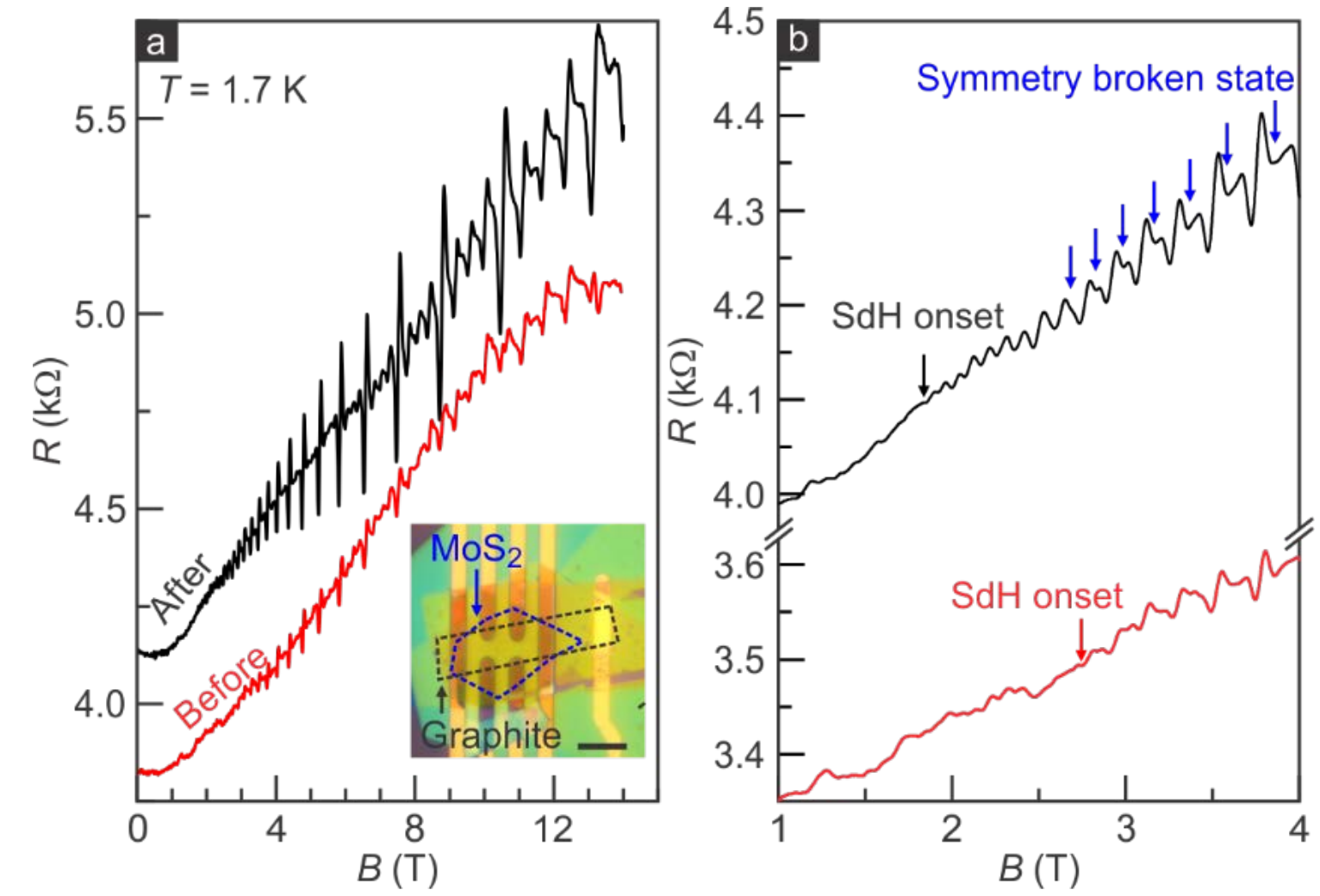

**Figure 2.** (a) Shubnikov-de Haas oscillations on device M1 with a density of $5 \times 10^{12}$ $cm^{-2}$ at $T$ = 1.7 K. Red and black lines show results before and after contact mode AFM, respectively. The red magnetotransport curve is vertically offset (-0.5 kΩ) for clarity. The inset displays an optical image of the device. The black dotted line delineates the graphite used as a top gate while the $MoS_2$ flake is indicated by the blue dotted line. The scale bar corresponds to 10 μm. (b) Magnified low magnetic field regime of (a). Black and red arrows mark the onset of Shubnikov-de Haas oscillations, whereas symmetry broken states are highlighted with blue arrows.

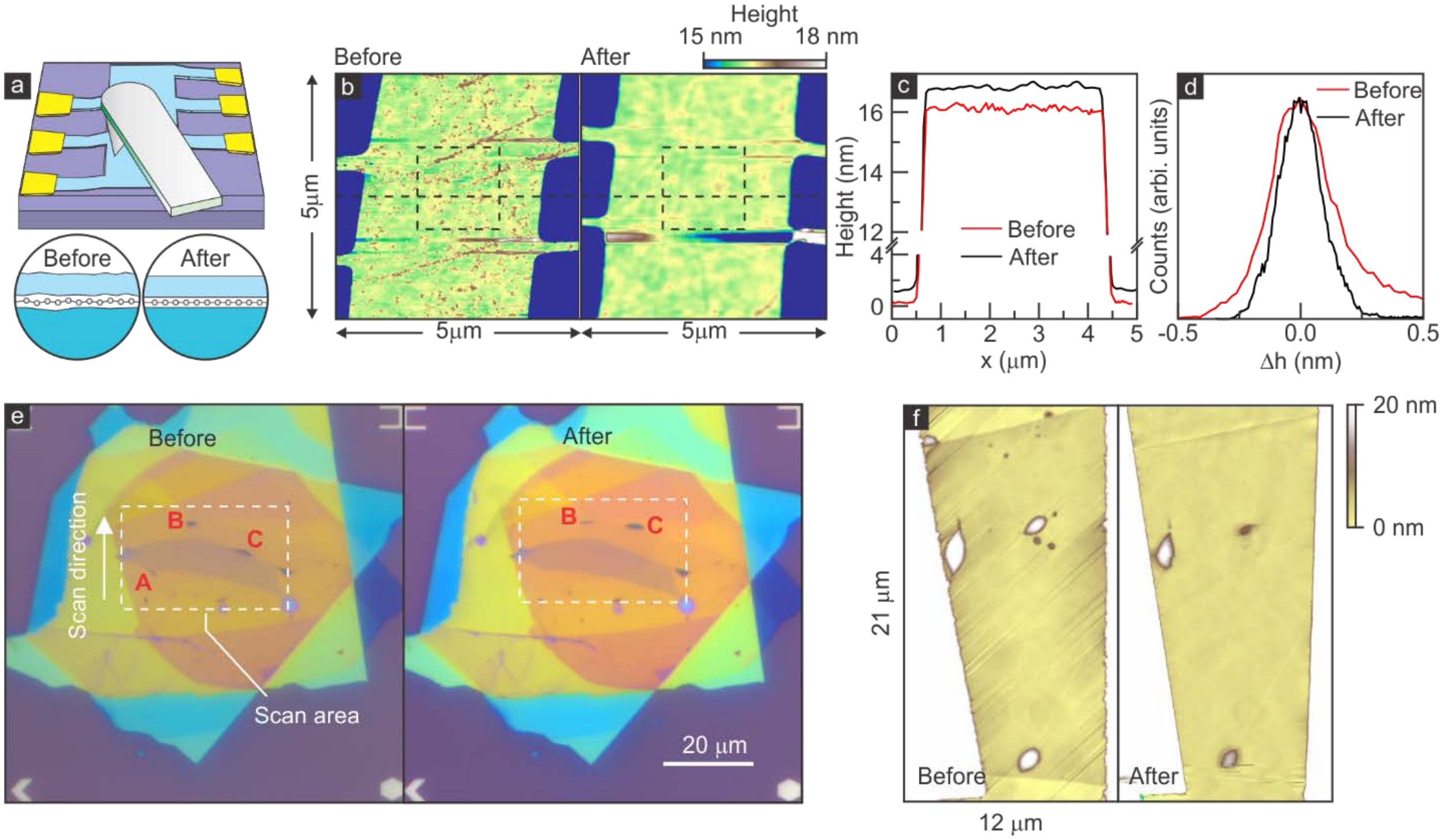

**Figure 3.** (a) Schematic of the van der Waals heterostructure device geometry and the AFM treatment. The bottom panels highlight the ironing effect. (b) Atomic force microscopy images of graphene based heterostructure G1 before (Left) and after (Right) AFM treatment. (c) Height profiles measured along the horizontal dashed lines in (b). The black curve is vertically offset with 1 nm for clarity. (d) Histogram of the height variation Δh in the areas demarcated by the boxes in (b). (e) Optical image of a van der Waals heterostructure before (left) and after (right) contact mode AFM. The heterostructure consists of a total of six 2D crystals: hBN, graphite, hBN, graphene, hBN, graphite (top to bottom). The area treated by the AFM as well as the scan direction have been imposed on the optical image. Three large bubbles are marked as A, B, and C. A has disappeared after the treatment, B has shrunk and C has moved. (f) AFM image recorded in non-contact mode of a hBN/$MoS_2$/hBN van der Waals heterostructure before (left) and after (right) contact mode AFM cleaning. White areas on the left and right side of the flake correspond to Cr/Au electrodes. Large bubbles have shrunk in size, smaller ones have disappeared after AFM cleaning.

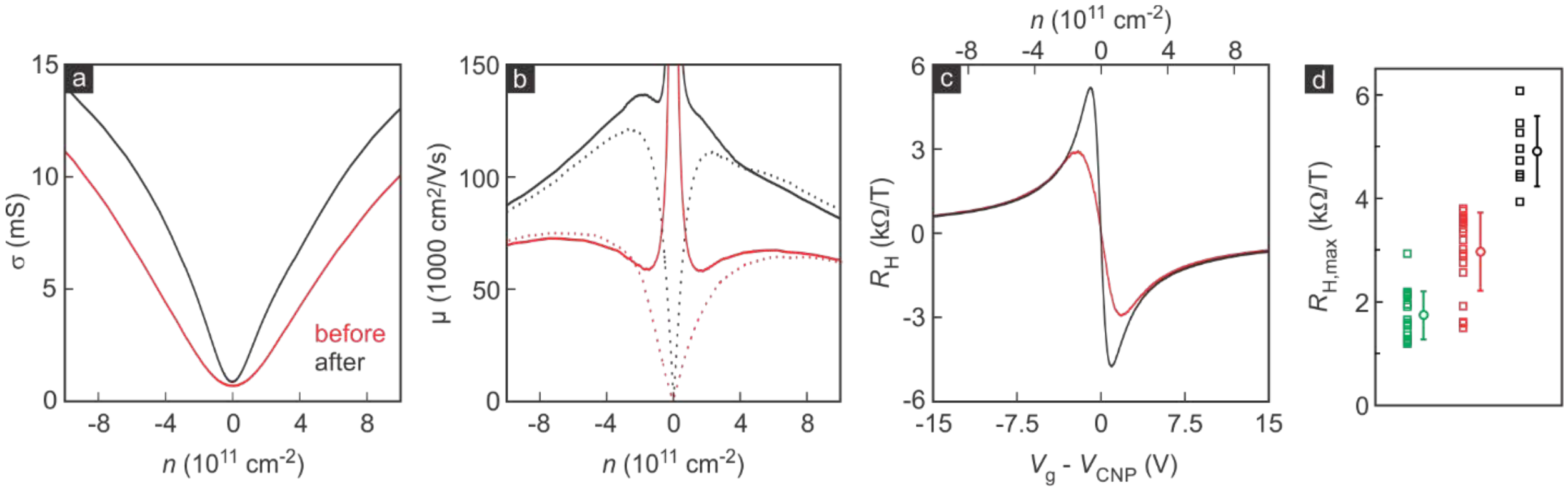

**Figure 4.** Room temperature transport measurements on graphene Hall bar device G2. (a) Conductivity as a function of gate controlled density before and after the AFM ironing step. The conductivity minimum at the CNP clearly narrows and the mobility increases. (b) Field effect mobility (solid lines) and Hall mobility (dotted lines) derived from the conductivity and Hall coefficient measurements. (c) Hall coefficient measured in weak magnetic fields as a function of gate voltage. The top axis shows the corresponding carrier density assuming a single carrier type. After the AFM treatment, substantially larger maximum Hall coefficients are obtained. These Hall coefficients correspond to the Hall sensor sensitivity. (d) Maximum Hall coefficients of a large ensemble of graphene devices. Each data point was recorded on a separate device. Black data points were obtained on hBN-encapsulated devices that were manufactured according to the full cleaning protocol, consisting of a 500°C annealing step before device processing and an AFM treatment afterwards. Red data points are hBN-encapsulated devices that were annealed but no contact AFM step was performed. For the devices shown, the annealing temperature before device processing ranged from 200°C up to 500°C. We found a slight improvement with increasing temperature but the impact is much smaller than the AFM treatment itself. Green data points correspond to uncovered graphene devices on $SiO_2$/Si substrates. They serve as reference. The mean values and standard deviations for each device type have been included as a circle with the error bar in the respective color.

## Associated Content

### Supporting Information

Increase of bubble free area by annealing, importance of 500°C anneal for significant quality enhancement upon AFM ironing, low temperature conductivity measurements and residual density estimates on additional graphene devices, and device statistics for key room temperature figure of merits using different processing protocols

### Author Information

Y. K. and P. H. contributed equally

### Corresponding Author

*E-mail: j.smet@fkf.mpg.de

### Author Contributions

Y.K., P.H., and J.H.S. conceived the project. Y. K. and P.H. carried out the device fabrication and performed the electrical measurements. T.T and K.W synthesized the h-BN crystal. Y.K., P.H, and J.H.S wrote the manuscript.

### Acknowledgements

We thank K. von Klitzing and D. S. Lee for fruitful discussions. We acknowledge Y. Stuhlhofer, S. Göres, J. Mürter, and M. Hagel for assistance with sample preparation. Y.K. is grateful for financial aid from the Alexander von Humboldt Foundation. The growth of hexagonal boron nitride crystals was sponsored by the Elemental Strategy Initiative conducted by the MEXT, Japan and the CREST (JPMJCR15F3), JST. J.H.S. acknowledges support from the graphene flagship (WP1).

## References

1. Novoselov, K. S.; Jiang, D.; Schedin, F.; Booth, T. J.; Khotkevich, V. V; Morozov, S. V; Geim, A. K. Two-Dimensional Atomic Crystals. *Proc. Natl. Acad. Sci.* **2005**, *102*, 10451–10453.

2. Du, X.; Skachko, I.; Barker, A.; Andrei, E. Y. Approaching Ballistic Transport in Suspended Graphene. *Nat. Nanotechnol.* **2008**, *3*, 491–495.

3. Bolotin, K. I.; Ghahari, F.; Shulman, M. D.; Stormer, H. L.; Kim, P. Observation of the Fractional Quantum Hall Effect in Graphene. *Nature* **2009**, *462*, 196–199.

4. Wang, L.; Meric, I.; Huang, P. Y.; Gao, Q.; Gao, Y.; Tran, H.; Taniguchi, T.; Watanabe, K.; Campos, L. M.; Muller, D. A.; Guo, J.; Kim, P.; Hone, J.; Shepard, K. L.; Dean, C. R. One-Dimensional Electrical Contact to a Two-Dimensional Material. *Science.* **2013**, *342*, 614–617.

5. Ki, D. K.; Fal'Ko, V. I.; Abanin, D. A.; Morpurgo, A. F. Observation of Even Denominator Fractional Quantum Hall Effect in Suspended Bilayer Graphene. *Nano Lett.* **2014**, *14*, 2135–2139.

6. Kim, Y.; Lee, D. S.; Jung, S.; Skákalová, V.; Taniguchi, T.; Watanabe, K.; Kim, J. S.; Smet, J. H. Fractional Quantum Hall States in Bilayer Graphene Probed by Transconductance Fluctuations. *Nano Lett.* **2015**, *15*, 7445–7451.

7. Li, J. I. A.; Tan, C.; Chen, S.; Zeng, Y.; Taniguchi, T.; Watanabe, K.; Hone, J.; Dean, C. R. Even-Denominator Fractional Quantum Hall States in Bilayer Graphene. *Science.* **2017**, *358*, 648–652.

8. Zibrov, A. A.; Kometter, C.; Zhou, H.; Spanton, E. M.; Taniguchi, T.; Watanabe, K.; Zaletel, M. P.; Young, A. F. Tunable Interacting Composite Fermion Phases in a Half-Filled Bilayer-Graphene Landau Level. *Nature* **2017**, *549*, 360–364.

9. Zibrov, A. A.; Spanton, E. M.; Zhou, H.; Kometter, C.; Taniguchi, T.; Watanabe, K.; Young, A. F. Even-Denominator Fractional Quantum Hall States at an Isospin Transition in Monolayer Graphene. *Nat. Phys.* **2018**, *14*, 930–935.

10. Cao, Y.; Fatemi, V.; Fang, S.; Watanabe, K.; Taniguchi, T.; Kaxiras, E.; Jarillo-Herrero, P. Magic-Angle Graphene Superlattices: A New Platform for Unconventional Superconductivity. *Nature.* **2018**, *556*, 43–50.

11. Kretinin, A. V.; Cao, Y.; Tu, J. S.; Yu, G. L.; Jalil, R.; Novoselov, K. S.; Haigh, S. J.; Gholinia, A.; Mishchenko, A.; Lozada, M.; Georgiou, T.; Woods, C. R.; Withers, F.; Blake, P.; Eda, G.; Wirsig, A.; Huco, C.; Watanabe, K.; Taniguchi, T.; Geim, A. K.; Gorbachev, R.V. Electronic Properties of Graphene Encapsulated with Different Two-Dimensional Atomic Crystals. *Nano Lett.* **2014**, *14*, 3270–3276.

12. Haigh, S. J.; Gholinia, A.; Jalil, R.; Romani, S.; Britnell, L.; Elias, D. C.; Novoselov, K. S.; Ponomarenko, L. A.; Geim, A. K.; Gorbachev, R. Cross-Sectional Imaging of Individual Layers and Buried Interfaces of Graphene-Based Heterostructures and Superlattices. *Nat. Mater.* **2012**, *11*, 764–767.

13. Pizzocchero, F.; Gammelgaard, L.; Jessen, B. S.; Caridad, J. M.; Wang, L.; Hone, J.; Bøggild, P.; Booth, T. J. The Hot Pick-Up Technique for Batch Assembly of Van der Waals Heterostructures. *Nat. Commun.* **2016**, *7*, 11894.

14. Purdie, D. G.; Pugno, N. M.; Taniguchi, T.; Watanabe, K.; Ferrari, A. C.; Lombardo, A. Cleaning Interfaces in Layered Materials Heterostructures. *Nat. Commun.* **2018**, *9*, 5387.

15. Uwanno, T.; Hattori, Y.; Taniguchi, T.; Watanabe, K.; Nagashio, K. Fully Dry PMMA Transfer of Graphene on h-BN Using a Heating/Cooling System. *2D Mater.* **2015**, *2*, 041002.

16. Dean, C. R.; Young, A. F.; Meric, I.; Lee, C.; Wang, L.; Sorgenfrei, S.; Watanabe, K.; Taniguchi, T.; Kim, P.; Shepard, K. L.; Hone, J. Boron Nitride Substrates for High-Quality Graphene Electronics. *Nat. Nanotechnol.* **2010**, *5*, 722–726.

17. Burson, K. M.; Cullen, W. G.; Adam, S.; Dean, C. R.; Watanabe, K.; Taniguchi, T.; Kim, P.; Fuhrer, M. S. Direct Imaging of Charged Impurity Density in Common Graphene Substrates. *Nano Lett.* **2013**, *13*, 3576–3580.

18. Amet, F.; Bestwick, A. J.; Williams, J. R.; Balicas, L.; Watanabe, K.; Taniguchi, T.; Goldhaber-Gordon, D. Composite Fermions and Broken Symmetries in Graphene. *Nat. Commun.* **2015**, *6*, 5838.

19. Couto, N. J. G.; Costanzo, D.; Engels, S.; Ki, D.-K.; Watanabe, K.; Taniguchi, T.; Stampfer, C.; Guinea, F.; Morpurgo, A. F. Random Strain Fluctuations as Dominant Disorder Source for High-Quality On-Substrate Graphene Devices. *Phys. Rev. X* **2014**, *4*, 041019.

20. Castellanos-Gomez, A.; Buscema, M.; Molenaar, R.; Singh, V.; Janssen, L.; Van Der Zant, H. S. J.; Steele, G. A. Deterministic Transfer of Two-Dimensional Materials by All-Dry Viscoelastic Stamping. *2D Mater.* **2014**, 011002.

21. Masubuchi, S.; Morimoto, M.; Morikawa, S.; Onodera, M.; Asakawa, Y.; Watanabe, K.; Taniguchi, T.; Machida, T. Autonomous Robotic Searching and Assembly of Two-Dimensional Crystals to Build van der Waals Superlattices. *Nat. Commun.* **2018**, *9*, 1413.

22. Kim, Y.; Herlinger, P.; Moon, P.; Koshino, M.; Taniguchi, T.; Watanabe, K.; Smet, J. H. Charge Inversion and Topological Phase Transition at a Twist Angle Induced van Hove Singularity of Bilayer Graphene. *Nano Lett.* **2016**, *16*, 5053–5059.

23. Kim, Y.; Balram, A. C.; Taniguchi, T.; Watanabe, K.; Jain, J. K.; Smet, J. H. Even Denominator Fractional Quantum Hall States in Higher Landau Levels of Graphene. *Nat. Phys.* **2019**, *15*, 154–158.

24. Ashcroft, N. W.; Mermin, N. D. Solid State Physics, Saunders College Publishing: Orlando. **1976**.

25. Datta, S. Electronic Transport in Mesoscopic Systems, Cambridge University Press: Cambridge. **1997**.

26. Goossens, A. M.; Calado, V. E.; Barreiro, A.; Watanabe, K.; Taniguchi, T.; Vandersypen, L. M. K. Mechanical Cleaning of Graphene. *Appl. Phys. Lett.* **2012**, *100*, 98–101.

27. Lindvall, N.; Kalabukhov, A.; Yurgens, A. Cleaning Graphene Using Atomic Force Microscope. *J. Appl. Phys.* **2012**, *111*, 064904.

28. Garcia, A. G. F.; Neumann, M.; Amet, F.; Williams, J. R.; Watanabe, K.; Taniguchi, T.; Goldhaber-Gordon, D. Effective Cleaning of Hexagonal Boron Nitride for Graphene Devices. *Nano Lett*. **2012**, 12, 4449-4454.

29. Miyazaki, H.; Odaka, S.; Sato, T.; Tanaka, S.; Goto, H.; Kanda, A.; Tsukagoshi, K.; Ootuka, Y.; Aoyagi, Y. Inter-Layer Screening Length to Electric Field in Thin Graphite Film. *Appl. Phys. Express.* **2008**, *1*, 034007.

30. de Juan, F.; Cortijo, A.; Vozmediano, M. A. H. Charge Inhomogeneities due to Smooth Ripples in Graphene Sheets. *Phys. Rev. B* **2007**, *76*, 165409.

31. Gibertini, M.; Tomadin, A.; Guinea, F.; Katsnelson, M. I.; Polini, M. Electron-Hole Puddles in the Absence of Charged Impurities. *Phys. Rev. B* **2012**, *85*, 201405(R).

32. Wang, L.; Gao, Y.; Wen, B.; Han, Z.; Taniguchi, T.; Watanabe, K.; Koshino, M.; Hone, J.; Dean, C. R. Evidence for a Fractional Fractal Quantum Hall Effect in Graphene Superlattices. *Science.* **2015**, *350*, 1231–1234.

33. Ribeiro-Palau, R.; Zhang, C.; Watanabe, K.; Taniguchi, T.; Hone, J.; Dean, C. R. Twistable Electronics with Dynamically Rotatable Heterostructures. *Science.* **2018**, *361*, 690–693.

34. Thomsen, J. D.; Gunst, T.; Gregersen, S. S.; Gammelgaard, L.; Jessen, B. S.; Mackenzie, D. M. A.; Watanabe, K.; Taniguchi, T.; Bøggild, P.; Booth, T. J. Suppression of Intrinsic Roughness in Encapsulated Graphene. *Phys. Rev. B* **2017**, *96*, 014101.

35. Rosenberger, M. R.; Chuang, H. J.; McCreary, K. M.; Hanbicki, A. T.; Sivaram, S. V.; Jonker, B. T. Nano-“Squeegee” for the Creation of Clean 2D Material Interfaces. *ACS Appl. Mater. Interfaces* **2018**, *10*, 10379-10387.

36. Wehrfritz, P.; Seyller, T. The Hall Coefficient: A Tool for Characterizing Graphene Field Effect Transistors. *2D Mater.* **2014**, *1*, 035004.

37. Chen, B.; Huang, L.; Ma, X.; Dong, L.; Zhang, Z.; Peng, L.-M. Exploration of Sensitivity Limit for Graphene Magnetic Sensors. *Carbon.* **2015**, *94*, 585–589.

38. Dauber, J.; Sagade, A. A.; Oellers, M.; Watanabe, K.; Taniguchi, T.; Neumaier, D.; Stampfer, C. Ultra-Sensitive Hall Sensors Based on Graphene Encapsulated in Hexagonal Boron Nitride. *Appl. Phys. Lett.* **2015**, *106*, 193501.

39. Joo, M. K.; Kim, J.; Park, J. H.; Nguyen, V. L.; Kim, K. K.; Lee, Y. H.; Suh, D. Large-Scale Graphene on Hexagonal-BN Hall Elements: Prediction of Sensor Performance without Magnetic Field. *ACS Nano* **2016**, *10*, 8803–8811.

40. Morozov, S. V.; Novoselov, K. S.; Katsnelson, M. I.; Schedin, F.; Elias, D. C.; Jaszczak, J. A.; Geim, A. K. Giant Intrinsic Carrier Mobilities in Graphene and Its Bilayer. *Phys. Rev. Lett.* **2008**, *100*, 016602.

41. Ho, D. Y. H.; Yudhistira, I.; Chakraborty, N.; Adam, S. Theoretical Determination of Hydrodynamic Window in Monolayer and Bilayer Graphene from Scattering Rates. *Phys. Rev. B* **2018**, *97*, 121404.

42. Mayorov, A. S.; Elias, D. C.; Mukhin, I. S.; Morozov, S. V.; Ponomarenko, L. A.; Novoselov, K. S.; Geim, A. K.; Gorbachev, R. V. How Close Can One Approach the Dirac Point in Graphene Experimentally? *Nano Lett.* **2012**, *12*, 4629–4634.

43. Movva, H. C. P.; Fallahazad, B.; Kim, K.; Larentis, S.; Taniguchi, T.; Watanabe, K.; Banerjee, S. K.; Tutuc, E. Density-Dependent Quantum Hall States and Zeeman Splitting in Monolayer and Bilayer $WSe_2$. *Phys. Rev. Lett.* **2017**, *118*, 247701

44. Larentis, S.; Movva, H. C. P.; Fallahazad, B.; Kim, K.; Behroozi, A.; Taniguchi, T.; Watanabe, K.; Banerjee, S. K.; Tutuc, E. Large effective mass and interaction-enhanced Zeeman splitting of K-valley electrons in $MoSe_2$. *Phys. Rev. B.* **2018,** *97*, 201407(R)

45. Pisoni, R.; Kormányos, A.; Brooks, M.; Lei, Z.; Back, P.; Eich, M.; Overweg, H.; Lee, Y.; Rickhaus, P.; Watanabe, K.; Taniguchi, T.; Imamoglu, A.; Burkard, G.; Ihn, T.; Ensslin, K. Interactions and Magnetotransport through Spin-Valley Coupled Landau Levels in Monolayer $MoS_2$. *Phys. Rev. Lett.* **2018**, *121*, 247701

46. Lee, J. E.; Ahn, G.; Shim, J.; Lee, Y. S.; Ryu, S. Optical Separation of Mechanical Strain from Charge Doping in Graphene. *Nat. Commun.* **2012**, *3*, 1024.

47. Banszerus, L.; Janssen, H.; Otto, M.; Epping, A.; Taniguchi, T.; Watanabe, K.; Beschoten, B.; Neumaier, D.; Stampfer, C. Identifying Suitable Substrates for High-Quality Graphene-Based Heterostructures. *2D Mater.* **2017**, *4*, 025030.

48. Neumann, C.; Reichardt, S.; Venezuela, P.; Drögeler, M.; Banszerus, L.; Schmitz, M.; Watanabe, K.; Taniguchi, T.; Mauri, F.; Beschoten, B.; Rotkin, S. V.; Stampfer, C. Raman Spectroscopy as Probe of Nanometre-Scale Strain Variations in Graphene. *Nat. Commun*. **2015**, *6*, 8429.

49. Ando, T.; Nakanishi, T.; Saito, R. Berry's Phase and Absence of Back Scattering in Carbon Nanotubes. *J. Phs. Soc. Jpn*. **1998**, 67, 2857-2862.

50. McEuen, P. L.; Bockrath, M.; Cobden, D. H.; Yoon, Y-G.; Louie, S. G.; Disorder, Pseudospins, and Backscattering in Carbon Nanotubes. *Phys. Rev. Lett.* **1999**, 83, 5098-5101

51. Amorima, B.; Cortijo, A.; de Juan,F.; Grushin, A.G.; Guinea, F.; Gutiérrez-Rubio, A.; Ochoaa, H.; Parente, V.; Roldán, R.; San-Jose, P.; Schiefele, J.; Sturla, M.; Vozmediano, M.A.H. Novel Effects of Strains in Graphene and Other Two Dimensional Materials. *Phys. Rep.* **2016**, 617, 1–54

52. Castro Neto, A. H.; Guinea, F.; Peres, N. M. R.; Novoselov, K. S.; Geim, A. K. The Electronic Properties of Graphene. *Rev. Mod. Phys*. **2009**, 81, 109-162

53. Sohier, T.; Calandra, M.; Park, C. H.; Bonini, N.; Marzari, N.; Mauri, F. Phonon-Limited Resistivity of Graphene by First-Principles Calculations: Electron-Phonon Interactions, Strain-Induced Gauge Field, and Boltzmann Equation. *Phys. Rev. B* **2014**, *90*, 125414.

54. Wiedmann, S.; Van Elferen, H. J.; Kurganova, E. V.; Katsnelson, M. I.; Giesbers, A. J. M.; Veligura, A.; Van Wees, B. J.; Gorbachev, R. V.; Novoselov, K. S.; Maan, J. C.; Zeitler, U. Coexistence of Electron and Hole Transport in Graphene. *Phys. Rev. B* **2011**, *84*, 115314.

55. Das Sarma, S.; Hwang, E. H.; Li, Q. Disorder by Order in Graphene. *Phys. Rev. B* **2012**, *85*, 195451.

56. Das Sarma, S.; Hwang, E. H Density-Dependent Electrical Conductivity in Suspended Graphene: Approaching the Dirac Point in Transport. *Phys. Rev. B* **2013**, *87*, 035415.

57. Ki, D. K.; Morpurgo, A. F. High-Quality Multiterminal Suspended Graphene Devices. *Nano Lett*. **2013**, *13*, 5165-5170.

58. Nam, Y.; Ki, D. K.; Soler-Delgado, D.; Morpurgo, A. F. Electron–Hole Collision Limited Transport in Charge-Neutral Bilayer Graphene. *Nat. Phys.* **2017**, *13*, 1207–1214.

**Supporting Information for**

**"Reliable Postprocessing Improvement of van der Waals Heterostructures"**

Youngwook Kim[†, ‡, ‖], Patrick Herlinger[†, ‖], Takashi Taniguchi[§], Kenji Watanabe[§], and Jurgen H. Smet[†,*]

[†]*Max-Planck-Institut für Festkörperforschung, 70569 Stuttgart, Germany*

[‡]*Department of Emerging Materials Science, DGIST, 42988, Daegu, Korea*

[§] *National Institute for Materials Science, 1-1 Namiki, Tsukuba, 305-0044, Japan*

[‖]These authors contributed equally

[*]E-mail: j.smet@fkf.mpg.de

S1. Increase of bubble free area by annealing

To illustrate the beneficial effect of thermal annealing, we provide dark field optical images of a hBN-graphene-hBN van der Waals heterostructure immediately after stacking as well as after two subsequent annealing steps. The first anneal is performed at 350°C and a second anneal proceeds at 500°C. With increasing annealing temperature, the bubbles merge and almost completely disappear during the final annealing step of 500°C. Non-contact AFM measurements also show that the remaining area is mostly free from bubbles (but not necessarily from sub-nanometer interfacial roughness that cannot be measured). Hence, the available area for designing the device is substantially increased. The graphene device G2, discussed in the main text, was manufactured from the stack shown here.

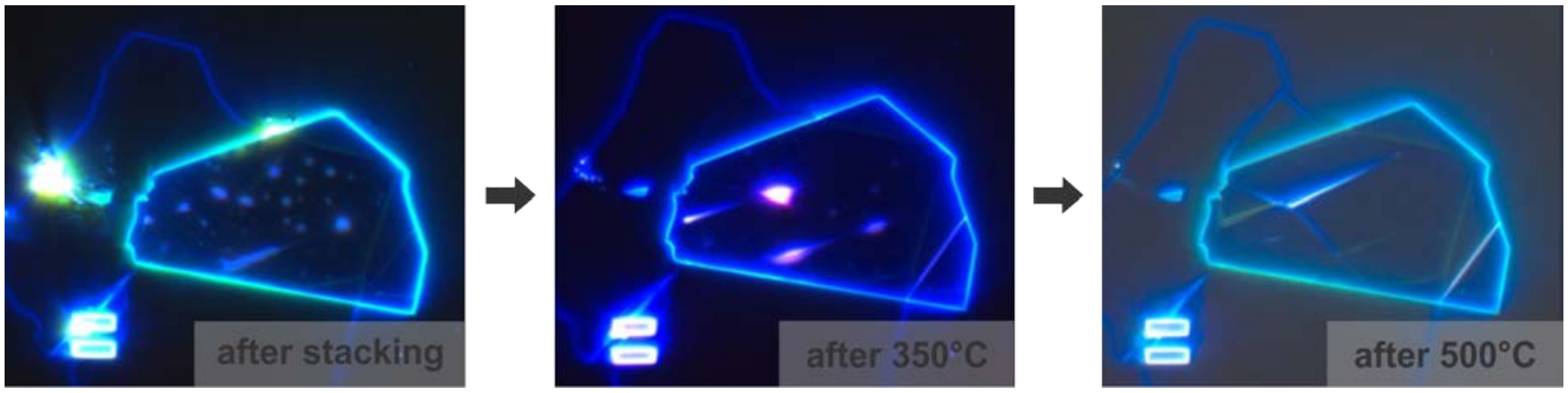

**Figure S1.** Dark field images of a hBN-graphene-hBN stack after fabrication and annealing (size of the marker consisting of two parallel lines at the bottom left is 5 µm). Light intensity in the right image, after 500°C annealing, was increased but no bubbles are visible. Instead, larger wrinkles

form mostly along the edges of the graphene flake. However, those wrinkles do not reduce the available area for the device.

## S2. Importance of 500°C anneal for significant quality enhancement upon AFM ironing

We find that for graphene as the active layer, an AFM treatment after the device fabrication has been completed only reliably causes a substantial improvement of the properties, if after stacking and prior to device fabrication the van der Waals heterostructure is annealed at 500°C. For the sake of completeness we note that we also conduct a final annealing step at 350°C after complete device processing, just prior to the AFM step. Figure S2 demonstrates exemplary the importance of the high temperature annealing step at 500°C after stacking for two devices. The same observations were also made on other devices.

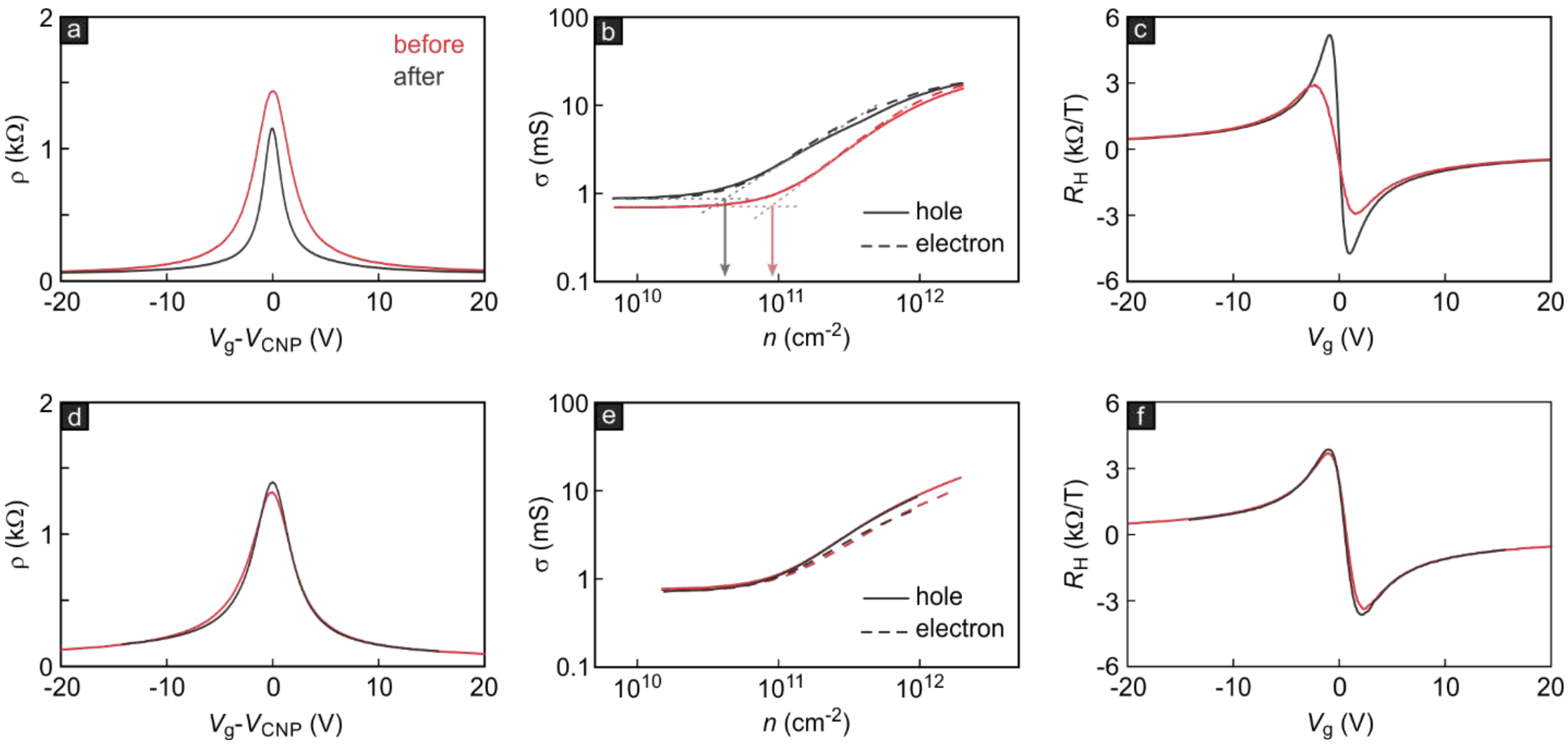

**Figure S2.** Room temperature magnetotransport measurements on two different hBN-encapsulated graphene devices before and after the AFM ironing step. Panel a-c: Resistivity, conductivity and Hall coefficient recorded on device G2. This device has been annealed at 500°C after stacking. It has already been discussed in Fig. 3 of the main text. Panels d-f: Same as for panel a-c but for a device that has been annealed only up to 350°C. For both devices another 350°C annealing step was performed after device processing. Only the first device shows a substantial improvement of the transport properties upon AFM ironing.

S3. Fabrication of devices with bottom electrodes

PMMA with an undercut profile has been used both as an etch mask as well as a self-aligned lift-off mask in order to produce recessed metallic contacts in a hBN flake. This procedure aims at reducing topography induced strain in $MoS_2$. An example of an AFM measurement on a sample with such recessed contacts is shown in Fig. S2. In this example the hBN is approximately 18 nm thick and about 21 nm of metal has been evaporated. We note that even though hBN may etch at a preferential angle, a protrusion with a height equal to the hBN thickness at the boundary of the metallic contact is not observed. The height increase or "ear" at the boundary of the contact is in the worst case about 0.7 nm high. We attribute the absence of larger protrusions at the contact boundary to the use of PMMA with undercut profile as a self-aligned etch and evaporation mask. Hence, with proper matching of the metal thickness, the height variation as we scan across the hBN and the metallic contact can be reduced substantially compared to the alternative approach of depositing the metallic contact directly on top of the hBN without prior etching, as a minimum contact thickness is required. We note for the sake of completeness that this alternative route of depositing the metal on top of the hBN without a recess has been pursued in Refs 43, 44 with contact thicknesses of either 9 or 12 nm, as well as in Ref 45. The latter publication does not contain metal thickness information.

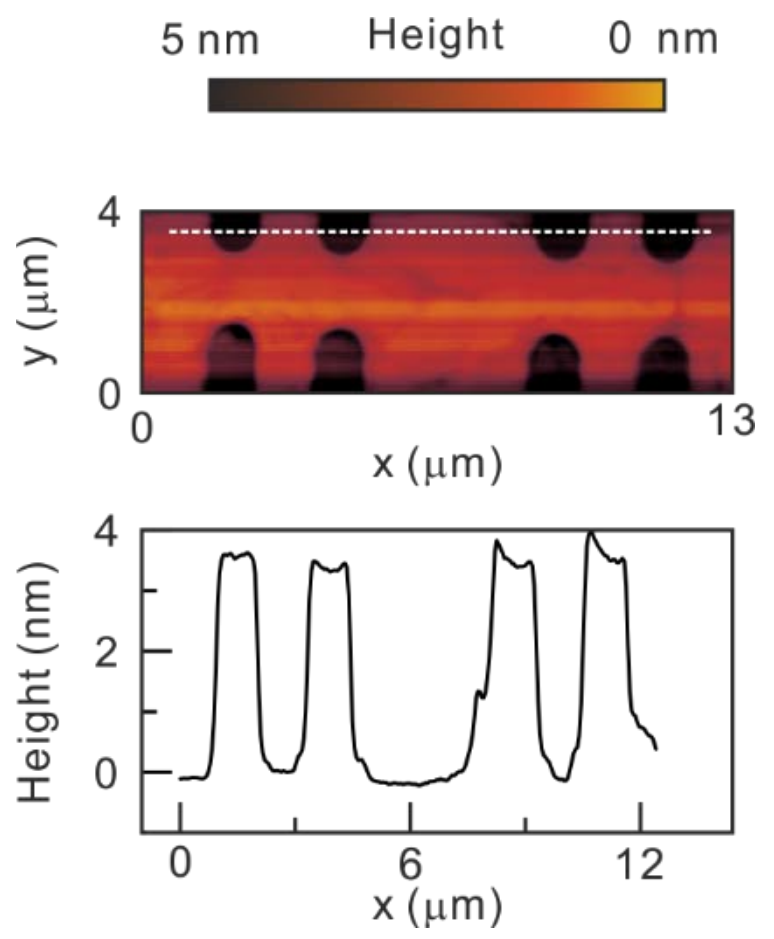

Fig. S3: AFM measurement of a sample in which metallic contacts are integrated within the bottom hBN layer. The deposited metal thickness is approximately 21 nm, whereas the hBN flake has a height of 18 nm. While there is a height increase at the contact boundary, it is well below 1 nm. In view of the geometry, the true topography is not hidden or affected by tip convolution.

S4. Raman spectroscopy statistics on devices produced with different fabrication protocols

In graphene's Raman spectrum, the G and 2D peak positions depend both on strain and doping level.[46] A disentanglement is at least partially possible when these peak positions are plotted in a 2D diagram. However, this relates to absolute strain or doping values. These may vary along a sample.[46,47] Hence, scanning along a stack will provide scattered data points within a certain region of the G/2D diagram. For each single measurement of the mapping process, Raman spectroscopy averages over the focal area on the scale of 1 µm. Since electron or hole puddles extend over a much smaller length scale, even though this scale is much larger than the sub-lattice spacing and hence long range, it is not possible to distinguish them by means of Raman spectroscopy independent of whether they may stem from charged impurities or nanometer strain fluctuations.

The averaging across the focal area does lead to a peak broadening and for instance Ref. 48 clearly attributes the 2D peak width to nanometer-scale strain variations. As also shown by Banszerus *et al.*[47] hBN-encapsulated graphene shows the narrowest 2D peak and the spread of the data across the sample is smaller than for other substrates. This already indicates the high quality of hBN-encapsulated stacks.

These publications report Raman maps on one and the same device or on very few devices to demonstrate the intra-sample spread. Here, single Raman spectra were recorded at a random position on each fabricated stack. Recording a full map on one and the same device would just reveal a cloud around the initial single data point as in the previous publications. Therefore, here we have chosen to compare the spectral characteristics recorded on different samples instead in order to obtain an estimate of the inter-sample spread. To the best of our knowledge, this quantity has, contrary to the intra-sample spread, not been addressed previously. The outcome of this elaborate study is summarized in Fig. S4 and S5.

As apparent from these figures, the G/2D lines as well as the 2D peak width are quite similar for all hBN-encapsulated devices fabricated by the now established pick-up process. Minor improvements in the 2D peak width and "scattering" of the G/2D peak position are achieved when the stacks are annealed at 500°C right after assembly but before device processing. This is our starting point for all devices in the present manuscript (pink open circles in Fig. S4). We find the data spread (scattering range) to remain the same after AFM treatment (hence not additionally shown). Such data points fell into the same data cloud. Hence, our experiments indicate that confocal Raman spectroscopy is not helpful in this particular case, even though it frequently is an extremely helpful and powerful characterization method. The transport quantities turn out more sensitive to the residual charge carrier density caused by strain fluctuations (the dominant source of disorder in clean devices), as shown in this work. This was already evident from Fig. 5 of the work by Couto and co-workers[19] where the mobility values for their hBN devices varied significantly, although the 2D peak width was always around 20/cm. The mobilities are inversely correlated to the residual charge carrier density. Hence, the changes in residual charge carrier density are not reflected in the 2D peak width in the high quality limit. Raman does of course help to discriminate bad and mediocre devices (hBN encapsulated without further treatment or without using the dry pick-up method) from good/excellent devices, but within the latter category it is difficult. Part of the problem is of course that Raman measurements are typically conducted at room temperature. For high quality devices, electron-hole puddles become invisible, because the thermally excited electron and holes outnumber the amount of charge carriers contained in the puddles and the averaging across the spot size makes graphene appear undoped.

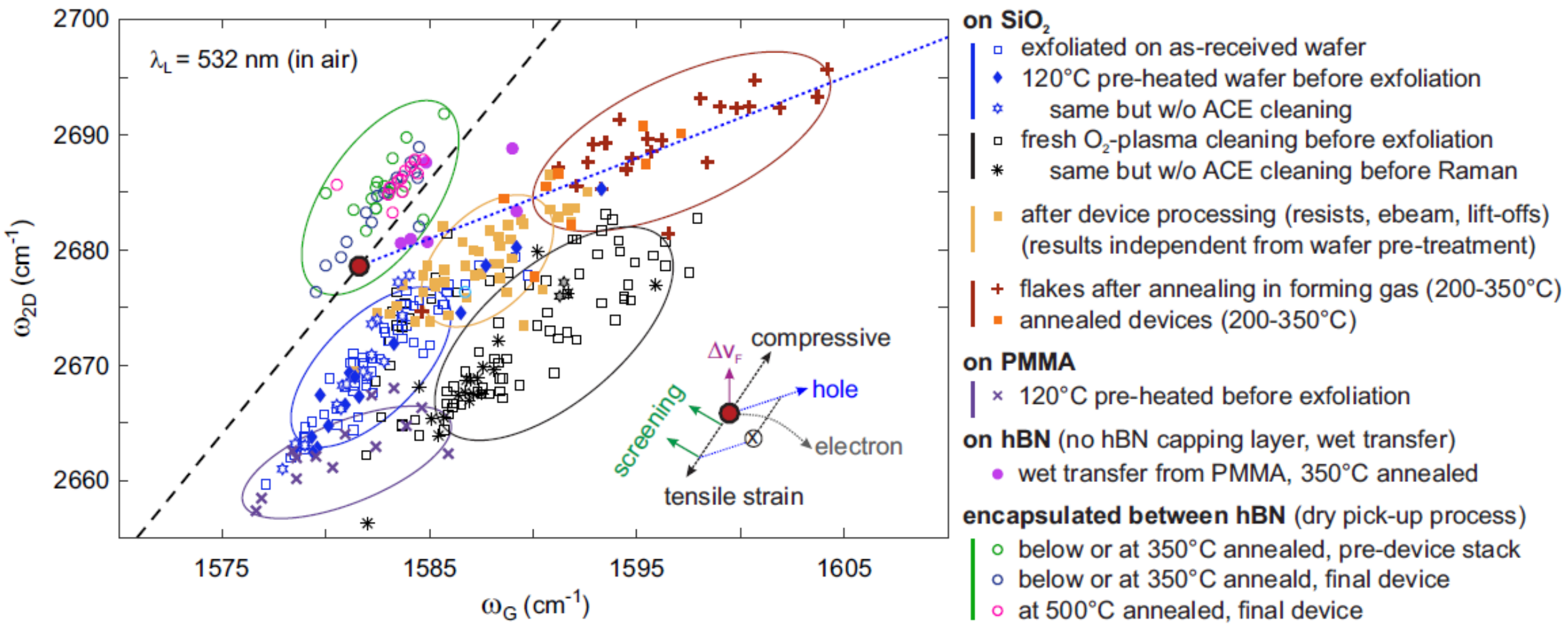

Fig. S4: Location of the 2D and G Raman lines for a large set of monolayer graphene samples. Each data point in this graph is coded by color and a symbol shape in order to distinguish whether the respective sample underwent any additional processing step as described in the legend, was encapsulated between hBN (open circles) or just supported on $SiO_2$, PMMA or hBN. The inset marks the expected movement of the data point when holes, electrons, tensile or compressive strain is added. Hence, the graph reveals insight in the impact of the substrate, encapsulation and processing on the absolute doping and strain level (averaged over the focal area). The data points can be grouped in six larger data sets. Each of them has been encircled by a thin colored line. The devices fully encapsulated between hBN using the dry pick-up method are located towards the top left of the graph inside the area encircled in green. They exhibit the least strain or doping in absolute terms. The annealing step at 500°C for these encapsulated devices slightly reduces the inter-sample spread. The additional AFM treatment on these samples (open pink circles), however, does not result in a change in the inter-sample spread (hence not shown here additionally). For $SiO_2$ as the substrate, one can distinguish between samples exfoliated onto as-received or mildly pre-heated wafers (blue area) and samples exfoliated onto plasma-treated wafers (black area). After device processing, strain and doping levels for both cases are typically found within the yellow area. Annealing of unprocessed or processed flakes on $SiO_2$ results in notable hole doping (red area), and hence should be avoided. Flakes exfoliated onto a pre-heated PMMA film (purple

area) are subject to substantial tensile strain. After wet transfer to hBN (without capping layer) this strain apparently releases.

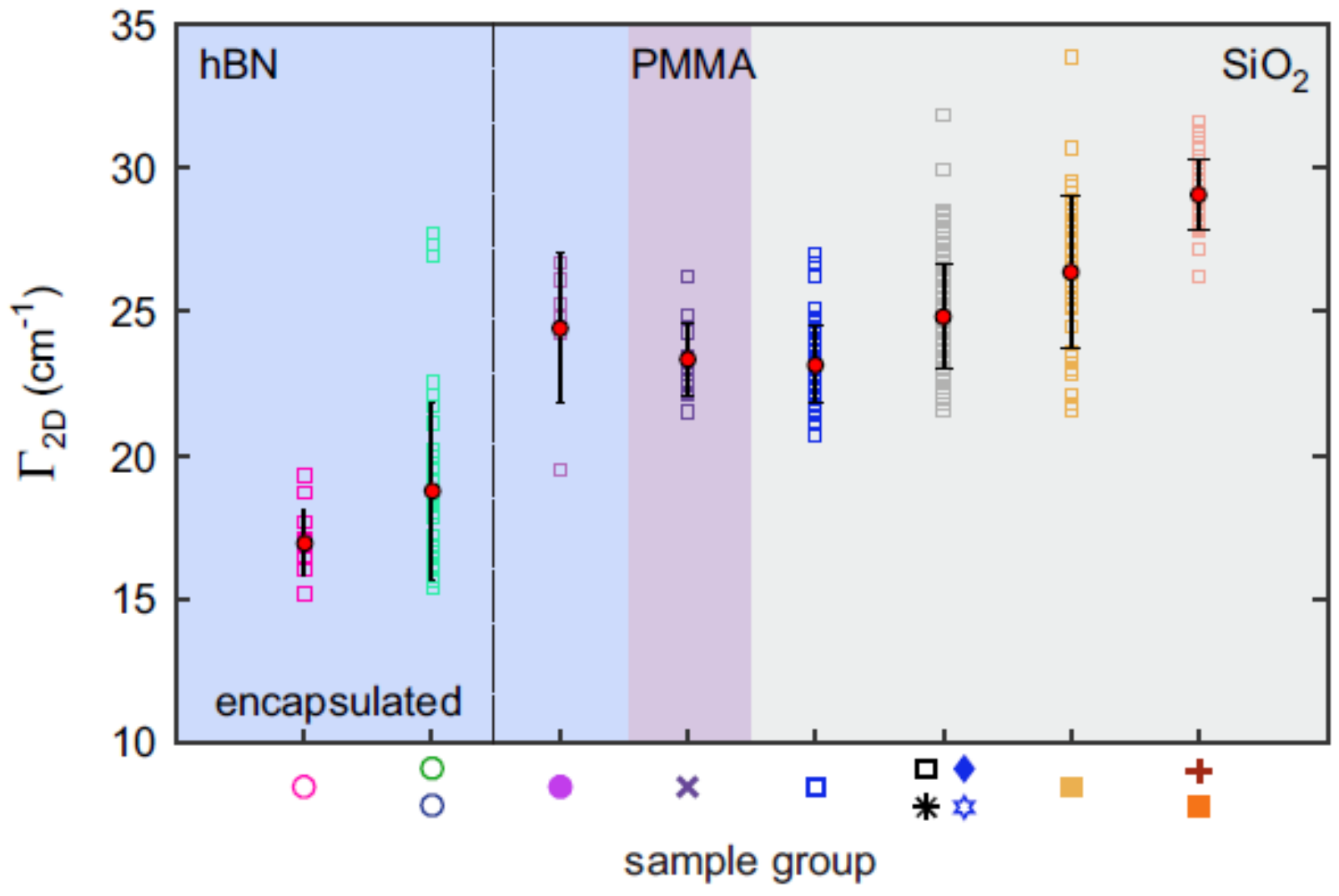

Fig. S5: Width of the 2D Raman line for all samples plotted in Fig. S4 (same legend applies). The width of the 2D band is a measure of the nanometer strain fluctuations (see Ref. 48). Starting from the right, the line width is the largest for samples on $SiO_2$ after annealing. In principal, all additional treatments after exfoliation on as-received wafers (blue open squares as reference) or wafers that have been pre-treated before exfoliation induces a slight increase of the 2D bandwidth. Only for the dry pick-up method the line width substantially shrinks and strain fluctuations are reduced. Among these encapsulated devices fabricated with this method, those that underwent the 500°C annealing step yield the best results. The subsequent AFM treatment causes no further improvement for the inter-sample spread for the reasons discussed in the text above.

S5. Impact of PMMA residue on top of the hBN cap layer

In the graph below, we show experimental data on a sample that was first treated by AFM ironing and then studied. Subsequently, it was intentionally contaminated with polymer residues and reinvestigated. The residues have no impact confirming the result of Wang *et al.*[4]

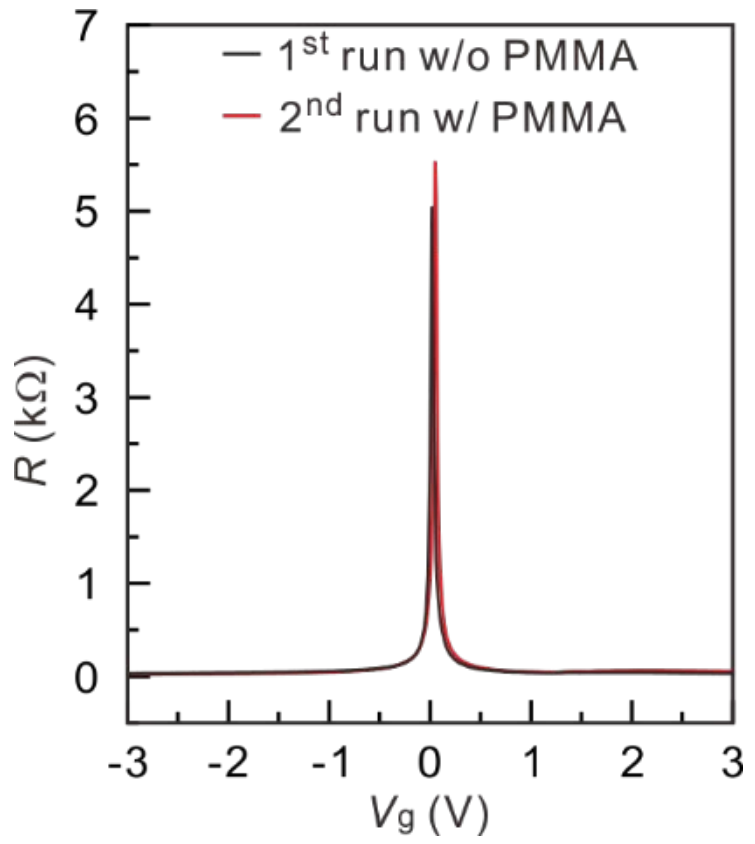

**Figure S6.** Gate dependence of the resistance recorded on a treated, cleaned sample and subsequently polymer “contaminated” sample. The sample has a graphite back gate with a thickness of 10 nm. The thickness of the top and bottom hBN is between 20 and 30 nm. The black curve is the resistance trace recorded after thermal annealing and a contact mode AFM treatment, while the resistance curve of the same device with polymer residues on the top surface is displayed in red. The Full-Width-Half-Maximum of both curves is 55mV (black) and 52 mV (red). The gate voltage difference at charge neutrality is only 0.003 V. This corresponds to a density difference at a given gate voltage of only 1 x $10^9$ $cm^{-2}$, below the accuracy with which we can determine the absolute value of the density.

S6. Discriminating different sources of density inhomogeneity

The large discrepancy between the scattering time, extracted from measured charge carrier mobilities, and the intervalley scattering time, obtained through an analysis of the weak localization effect, has allowed the authors of Ref. 19 to conclude that the disorder in graphene samples supported on different substrates is of long range character, in particular also for hBN substrates of interest here. Long range disorder either originates from charged impurities or from strain fluctuations. In practice, both sources are bound to contribute to disorder and need to be addressed in order to achieve the highest possible quality. For high quality, state-of-the-art samples it is however possible to discriminate which source is predominant. While both sources of disorder generate density inhomogeneity, only in the case of strain fluctuations there is an inverse correlation between the amplitude of density inhomogeneity and the transport mobility at low temperature, when phonon contributions and thermal activation of carriers can be neglected. How different transport properties are affected by both sources of disorder is discussed in more detail below.

The addition of graphitic gates has been an important advance in the field in recent years as it has reduced the density inhomogeneity in already clean encapsulated graphene devices by screening the charged impurities located in for instance the substrate or in residues or adsorbates at its surface.[8,18] These charged impurities generate a long-range disorder potential and contribute "amplitude" to the electron-hole puddle landscape. If one charge polarity dominates, it will also cause an overall shift in the charge neutrality point. In the presence of a magnetic field, the density inhomogeneity implies a local variation of the filling factor and therefore will also reduce the observability of quantum oscillations and the quantum Hall effect due to inhomogeneous broadening. In the absence of a magnetic field, the additional residual carrier density at overall charge neutrality will increase the minimal conductivity, i.e. the value at which the conductivity saturates as one approaches overall charge neutrality in the log-log presentation. However - and this is the crux - the conductivity of these initially already clean devices is not modified because of increased scattering induced by the long range disorder potential from remote charged impurities. Graphene has built in protection against backscattering induced by such a potential as the band structure is composed of two intersecting Bloch bands associated with the two sub-lattices (pseudospin).[19,49,50] States belonging to these two Bloch bands are orthogonal (opposite pseudospin) and a scalar potential can only cause a transition between states of both bands if it is short-ranged, i.e. if it varies on the scale of the sub-lattice or the C-C distance which equals 0.14 nm.

Local strain fluctuations also produce an inhomogeneous density landscape and contribute to the amplitude of the electron-hole puddles, however in addition they generate a spatially varying vector potential or effective gauge field.[19,51] While the density inhomogeneity induced by strain fluctuations affects the conductivity and the observability of quantum oscillations in an identical manner as when this density inhomogeneity would originate from charged impurities, the effective fluctuating gauge field has no analogue for the case of long-rang disorder induced by charge impurities. This gauge field is detrimental for the conductivity, since it does enable backscattering even within a single valley and degrades mobility, despite its long-range nature. This becomes apparent from a simple gedankenexperiment, sketched also in Ref. 52, on a graphene flake that consists of two regions: a region with no strain and a region where all horizontal bonds are compressed. In k-space the compression causes a change of the distance of the inequivalent Dirac cones. An electron, that initially travels horizontally in the unstrained region and crosses the boundary between these two regions, retains the same energy as well as its momentum in the direction along the boundary. Hence it remains on the same constant energy contour, but its velocity (which is always oriented perpendicular to the constant energy contour) no longer points along the horizontal direction once the boundary is crossed due to the displaced Dirac cones. Hence, the electrons get deflected. The weak localization studies conducted by Couto *et al.* (Ref. 19) enabled the extraction of the intravalley scattering time, which in their notation is the characteristic time to break the effective single-valley time-reversal symmetry $\tau^*$. Because this time was found to be comparable with the elastic scattering time obtained from the mobility values and since this intravalley scattering cannot occur with the assistance of a scalar long-range disorder potential induced by charged impurities – which is unable to break the single-valley time-reversal symmetry – it was concluded that the time scale for charge carrier scattering is a result of strain induced scattering according to the mechanism outlined above.

In summary, strain fluctuations and a long range disorder potential from remote charged impurities both increase the amplitude of the density inhomogeneity and affect quantum oscillations through inhomogeneous broadening. However, the strain fluctuations also severely degrade mobility. The inverse correlation between the density inhomogeneity and the mobility as observed in this work and the work of Couto *et al.* (Ref. 19) is a unique characteristic of strain induced disorder. This correlation represents unequivocal evidence for the importance of strain fluctuations in already clean devices.

For the sake of completeness, we note that residues are bound to be responsible not only for long-range charged impurity disorder, but also for strain fluctuations. For instances, differences in the thermal expansion coefficients will inevitably turn these residues into local stressors. As a result, removal of these residues will not only reduce the level of charged impurity disorder, but simultaneously strain induced disorder. Since both charged impurities as well as strain fluctuations can in principle produce electron-hole puddles, both can be responsible for a shift of the charge neutrality point. Observing an overall shift in the charge neutrality point can therefore not be invoked as a proof for the reduction of charged impurity disorder. The origin of shift can be either strain release or the reduction of charge impurity disorder.

Apart from the above arguments, our working hypothesis, that an ironing-like effect is primarily responsible for the boost in transport mobility and density homogeneity rather than remote charged impurities, also follows from the following findings:

- Contaminating intentionally the top hBN surface with polymer residues after the AFM cleaning does not show an effect on device quality.
- Bubbles at the interface can indeed be manipulated. Hence a manipulation of nanometer sized strain-fluctuations is very plausible.
- At low temperature our devices fall on the $\mu/n^*$ curve predicted by the theoretical model of random strain fluctuations as outlined for instance in Ref. 19. This correlation is not expected for charged impurity scattering.
- For a $MoS_2$ device equipped with both a bottom as well as a top graphitic gate to shield remote charged impurities, we demonstrate that our method still leads to a notable improvement.

S7. Low temperature conductivity measurements and residual density estimates on additional graphene devices

Figure S7 plots the conductivity as a function of density on a double log scale at 1.3 K recorded before and after contact mode AFM on three graphene devices (device G1, G3, and G4). Device G1 is the same device as discussed in the main text. The residual density typically drops down to $10^{10}$ $cm^{-2}$ or below consistent with the picture that random strain fluctuations are reduced by the AFM treatment.

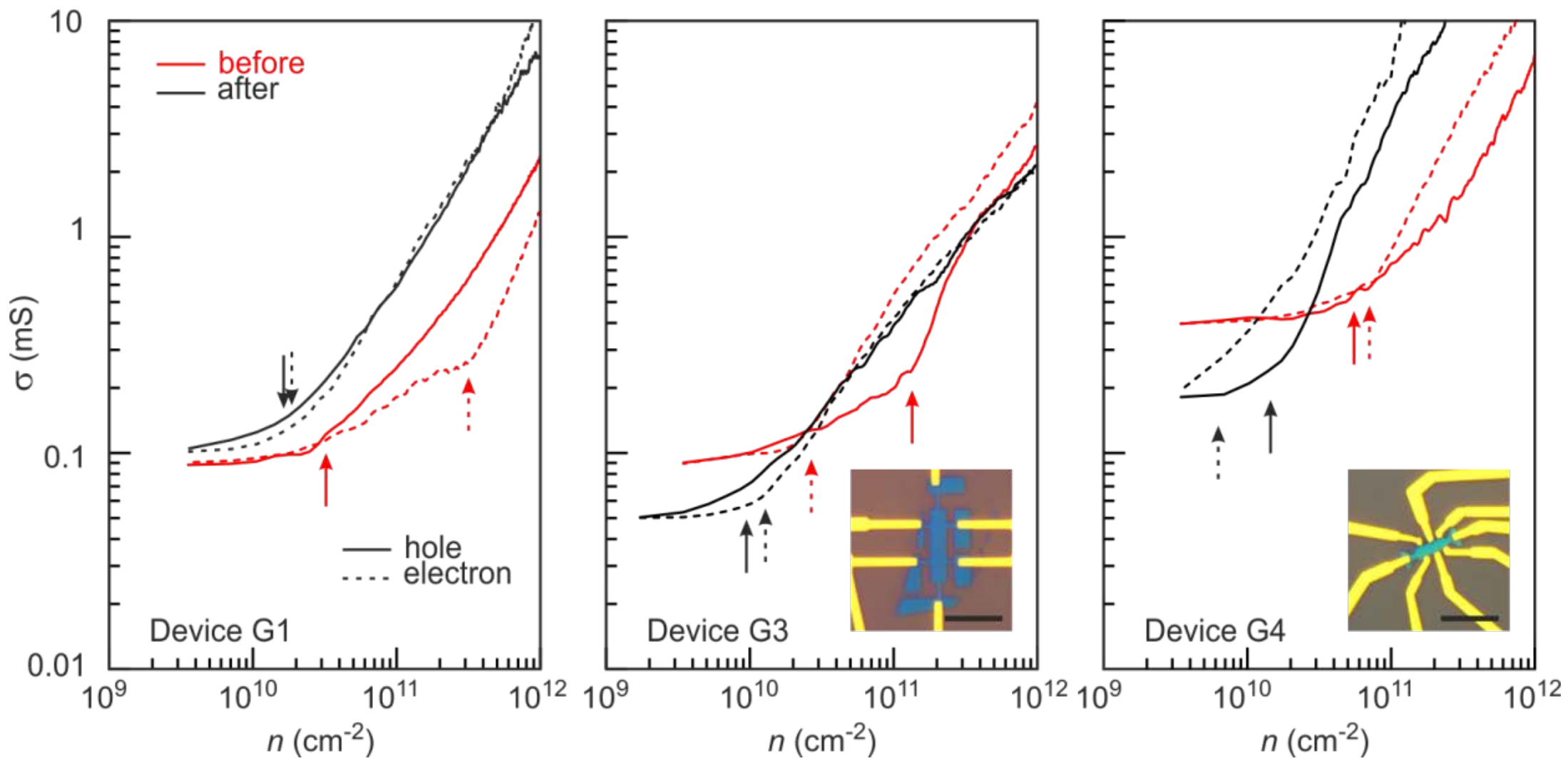

**Figure S7.** Conductivity as a function of density on a double logarithmic scale for device G1 (Left), G3 (middle) and G4 (right). These data have been recorded at 1.3 K. Red and black lines are measurements before and after contact mode AFM, respectively. Dotted lines are for electron transport and solid lines for hole transport. Arrows mark the estimated residual charge carrier density as determined from the crossing of a linear extrapolation of the conductivity at high density and the conductivity saturation at low density. Device images are displayed in the inset of each panel. The scale bars correspond to 10 μm scale bar. The image for device G1 can be found in panel a of Fig. 1.

S8. Device statistics for key room temperature figure of merits using different processing protocols

We compare key transport characteristics for different device fabrication protocols and demonstrate that the combination of a 500°C anneal after stacking and a post-processing AFM treatment reliably yields the best results. Panels a through e of Fig. S8 display the resistivity maximum at the CNP, the residual density at the CNP as estimated from a log-log plot of the conductivity *versus* density, the resistivity at a high density of $n = 10^{12}$ $cm^{-2}$ where only one carrier type is present and the mobility at this density, respectively. Each data point in every panel of Figure S8 corresponds to an individual device. Data points have been grouped into three columns. The first column reports data for devices fabricated out of bare graphene supported on $SiO_2$ without any additional treatment. These green data points serve as reference. All other devices consist of graphene flakes encapsulated in between two hBN flakes. Data on devices that were annealed at a maximum temperature of 350°C without conducting the final AFM treatment were plotted in the second column. Finally, the black data points in the third column were gathered on encapsulated graphene that underwent a 500°C anneal after stacking and AFM ironing after device processing. In all cases, the doped silicon substrate served as the back gate for tuning the carrier density. Figure S8 clearly shows that the transport properties are significantly and reliably enhanced using the fabrication and AFM treatment method outlined before. This becomes particularly apparent when adding an average of the data points and their standard deviation for each plotted quantity and processing protocol (square symbol and error bar). The variation among the data points has shrunk significantly for AFM treated samples. As discussed in Section S2, the AFM ironing does not lead to a notable improvement of the sample characteristics, if a van der Waals heterostructure is only annealed up to 350°C or less (this case is not shown in Fig. S8). We emphasize that the mobility values provided in Figure S8 are either calculated directly from the measured resistivity in the high density limit, where the single carrier picture clearly applies and the mobility values after AFM treatment approach the acoustic phonon limit, or are given as an estimate of the mobility at the CNP $\mu_{CNP} = 1 / (e\, n_{t,0}\, \rho_{max})$, where $\rho_{max}$ is the measured peak resistivity and $n_{t,0}$ the residual total carrier density at the CNP. The latter is obtained from independent Hall curve measurements and fitting. Typically, these $\mu_{CNP}$ values are notably smaller than the peak mobility values, since we consider here the presence of two carrier types. When comparing with existing literature, it should be kept in mind that there frequently only peak mobilities are quoted instead,[14] which are typically larger. Extracting peak mobility values for our devices, that means the maximum value of the field effect mobility defined as $\mu = 1 / e\, n\, \rho$ (using the single carrier density n), we obtain comparable values with those stated in Ref. 14 (see for instance Figure 4b in the main text).

The dashed line in panel b marks the thermally excited total density at room temperature (300 K) assuming a linear energy dispersion with Fermi velocity $v_F = 10^6$ m/s. It can be calculated analytically and reads $n_{t,0}(T) = (\pi\, k_B^2 / 3\, \text{hbar}^2\, v_F^2)\, T^2$.[41, 42] The experimentally obtained values are substantially smaller confirming previous studies that $v_F$ is larger in high quality devices due to renormalization effects (see also Ref. 42). The acoustic phonon scattering limits[4,53] have been included in panels d and e as grey shaded areas. At high carrier density this acoustic phonon limit is approached. Hence, we conclude that the post-processing AFM cleaned devices reach the intrinsic limit. There is also a clear correlation between the drop in the residual carrier density for the high quality AFM treated devices and the measured maximum in the Hall coefficient, proving the relation $R_{H,max} = 1 / 2 e n_{t,0}$ (see Figure 3d in the main text). Finally, supported by the statistics we have collected here, we are able to stipulate empirical criteria to identify high quality devices at room temperature based on the measurement of two sample dimension independent quantities: the resistivity at the CNP and the resistivity in the high density limit (here $n = 10^{12}$ $cm^{-2}$). The peak resistivity at the CNP should be around 1kΩ and the resistivity at high density below 100 Ω, better 50-70 Ω, or as a rule of thumb at least 1/10 of the peak resistivity. We note for the sake of completeness that the peak resistivity at the charge neutrality point always reflects transport in the diffusive regime and not the ballistic regime, irrespective of the sample quality.[41,54-58] Long range disorder dominates in the low temperature limit and obscures intrinsic Dirac cone physics. If temperature increases, optical phonon modes play a role at low density near the charge neutrality point. There may be a small window at intermediate temperatures where electron-electron scattering is important. It is mainly electron-hole scattering in case of the charge neutrality point. However, even this scattering is a diffusive process. We refer to the work by Ho *et al.* [41] and in particular Fig. 1 where the different regimes have been discussed at length.

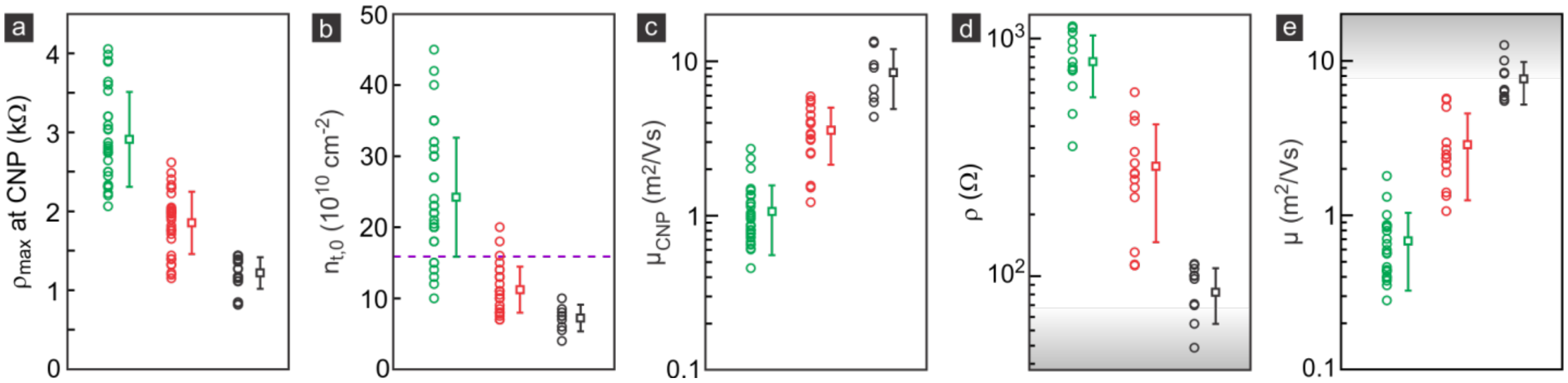

**Figure S8.** Statistics of the electrical transport properties of bare graphene and encapsulated graphene devices measured at room temperature. We distinguish between bare graphene devices on $SiO_2$ (green), hBN-encapsulated graphene devices annealed at 350°C without AFM treatment (red), as well as encapsulated devices which were annealed at 500°C after stacking and treated with AFM after the device processing has been completed (black data points). The data points for these three sets of devices are organized in three columns in each panel. Squares with error bars represent the average value of the plotted quantity and its standard deviation. (a) Maximum resistivity at the CNP as extracted from the resistivity curve as a function of the applied gate voltage. (b) Residual total carrier density at the CNP, as estimated from the double-logarithmic plot of the conductivity as a function of density (see panel b of Figure S2). The dashed line marks the thermal limit for a linear band structure dispersion with $v_F = 10^6$ m/s. (c) Estimate of the mobility at the CNP, using $\mu_{CNP} = 1 / (e\, n_{t,0}\, \rho_{max})$. (d) Resistivity at a carrier density of $n = 10^{12}$

cm$^{-2}$. At this density only one carrier type is present. (e) Mobility at n = $10^{12}$ cm$^{-2}$, derived from the resistivity value. Gray shaded areas in panels d and e mark the limit due to acoustic phonon scattering.[4,53] Note the logarithmic scale in panels c to e. The gate voltage to density conversion factors were determined from separate Hall measurements on each device.